\documentstyle[12pt,epsf]{article} 
\newcommand{\beq}{\begin{eqnarray}}
\newcommand{\eeq}{\end{eqnarray}}

\newcommand{\g}{\gamma}

\newcommand{\rf}[1]{(\ref{#1})}

\renewcommand{\b}{\beta}
\renewcommand{\a}{\alpha}

\newcommand{\th}{\theta}

\newcommand{\r}{\prime}

\newcommand{\WP}{{\cal P}}
\newcommand{\EE}{{\cal E}}
\newcommand{\ZZ}{{\cal Z}}

\begin{document}

\topmargin 0pt
\oddsidemargin 5mm
\headheight 0pt

\topskip 5mm


\thispagestyle{empty}
      
\begin{flushright}
NBI-HE-98-41\\
December 1998
\hfill
\end{flushright}

\begin{center}

\hspace{10cm}

\vspace{15pt}
{\large \bf
BARRIER PENETRATION IN $1+1$-DIMENSIONAL\\
$O(n)$ SIGMA MODELS }

\vspace{20pt}

{\bf Enrique Moreno}\footnote{\noindent Work supported by CUNY-Collaborative
grant 991999.}\\

The City University of New York,\\
C.C.N.Y. and Baruch College, New York, New York\\
moreno@scisun.sci.ccny.cuny.edu

\vspace{10pt}
and
\vspace{10pt}

{\bf Peter Orland}$^{1,}$\footnote{\noindent Work supported 
by PSC-CUNY Research Award Program 
grant
no. 668460.\\
$\dagger$Permanent Address}

The Niels Bohr Institute,\\
Blegdamsvej 17, DK-2100,
Copenhagen {\O}, Denmark\\
orland@alf.nbi.dk

\vspace{5pt}

and\\

\vspace{5pt}

The City University of New York,\\
Baruch College
and
The Graduate School and University Center,
New York, NY$^{\dagger}$ \\
orland@gursey.baruch.cuny.edu
\\

\vspace{10pt}

{\bf Abstract}

\end{center}

The $O(n)$ nonlinear sigma model 
in $1+1$ dimensions is examined as quantum mechanics
on an infinite-dimensional  
configuration space. Two metrics
are defined in this space. One of these metrics is
the same as Feynman's distance, but we
show his conclusions concerning 
potential energy versus distance 
from the classical vacuum are 
incorrect. The potential-energy functional 
is found to 
have 
barriers; the configurations on these
barriers are solitons of an associated sigma model
with an external source. The tunneling 
amplitude is computed 
for the $O(2)$ 
model and
soliton condensation is shown to drive 
the phase
transition
at a critical
coupling. We find the
tunneling
paths in the configuration space of the $O(3)$ model and argue
that these are responsible for the mass gap at $\theta=0$. These 
tunneling 
paths have half-integer topological charge, supporting
the conjecture due to Affleck and Haldane 
that there is a massless phase at
weak coupling and 
$\th=\pi$.

\vfill
\newpage
\pagestyle{plain}
\setcounter{page}{1}

\section{Introduction}

\setcounter{equation}{0}
\renewcommand{\theequation}{1.\arabic{equation}}

The two-dimensional $O(3)$ nonlinear sigma model has been regarded as interesting by
elementary-particle physicists because of the features it shares with QCD, such
as asymptotic freedom 
\cite{AF} and instantons 
\cite{INST}. Condensed-matter physicists have studied
the model in connection with antiferromagnetic spin chains \cite{SC}, 
the quantum hall effect \cite{QHE} and (more traditionally) 
classical ferromagnets. The
$O(n)$ model can be studied by the $1/n$ expansion and the $O(2)$ model can be understood 
a Coulomb gas of vortices \cite{kt}. The spectrum and S-matrix of the $O(3)$ 
model can be determined exactly \cite{ZZ}, \cite{PW}. Haldane argued that at $\theta=\pi$ the model should be
equivalent to the half-integer $XXX$ spin chain and is therefore gapless \cite{SC}.

In spite of these successes, there are
still unresolved issues. The 
spin-chain arguments, though
extremely
compelling, need to be supplemented by other 
evidence, since the ``Haldane mapping" \cite{SC} is 
not strictly a mapping at all, but
relies on some assumptions about local properties of the
ground state. Indeed, naively one might conclude from the spin-chain arguments that
any $O(3)$ invariant, integer-spin antiferromagnetic chain has a gap; yet
some integer-spin chains are 
gapless \cite{FadTakd}. 

In 
support of Haldane's conclusion 
Affleck and Haldane \cite{AffHal} 
suggested that
condensation of merons \cite{gross}, objects
with half-integer topological charge, was responsible for the gap at $\theta=0$, while
at $\theta=\pi$ and small bare coupling
the merons
bind in pairs, rendering the spectrum gapless. These arguments were not
precise for the $O(3)$ sigma model, but were made
by a conjectured extrapolation
from a model with $O(3)$ broken to $O(2)$. Nonetheless, numerical work of Bietenholz et. al.
with the symmetry unbroken suggests that this
picture is valid \cite{WPW}. Other observations concerning
the
role of half-integer topological charge were made recently in reference \cite{asorey}.

In this paper we study the $O(2)$ and $O(3)$ nonlinear sigma models 
using some ideas developed
for non-Abelian gauge theories \cite{orland}. Some
clarification of these ideas in reference \cite{orland}, as well as some preliminary work
for the sigma model is presented in reference \cite{kmo}.

The starting point is the definition of 
distance functionals on the fields. We work with two such functionals. One of these is the
natural distance related to the kinetic term. The other is defined on the orbit
space, which consists of
configurations modulo global
rotations. The latter distance functional 
is implicit in the work
of Feynman \cite{feynman}. Each of these functionals is a metric. The 
physical utility of these metrics are discussed. The second metric
is used to define a coordinate on the orbit space. We then
examine the minima of the potential energy for fixed values of this coordinate. What
appears is a structure of energy barriers in configuration space. On these barriers, the
configurations are identical to solitons - not of the ordinary sigma model, but
of a related classical field theory. These solitons are topologically stable for the $O(2)$
model, but not for the $O(3)$ model. They can be written down explicitly
in terms of elliptic and related functions. We show
that the barriers of the $O(2)$ model 
can be penetrated only above a critical coupling. At this
coupling a phase transition is driven by condensation of solitons. We thereby
obtain an entirely 
quantum-mechanical explanation of the transition, complementing the standard
statistical-mechanical explanation \cite{kt}. The barriers of the $O(3)$ sigma model
are considerably more complicated than
those of the $O(2)$ model. We find these and examine them in some detail. We
then show that the barrier-penetration process has topological 
charge one half; a result which strongly supports the arguments
of Affleck and Haldane \cite{AffHal}.

Our Hamiltonian analysis is quite different from the usual path-integral
saddle-point approach to the sigma model. The viewpoint
is one almost never considered for the model, except
in the appendix in Feynman's paper \cite{feynman}. As the style of his
article is 
difficult to follow, few physicists have
pursued the ideas further. We therefore see the need for
a nontechnical summary of our motivation and results, which is the
content of the next section.

\section{Distance and potential-energy topography} 

\setcounter{equation}{0}
\renewcommand{\theequation}{2.\arabic{equation}}

Field theories are infinite-dimensional quantum systems. In the Schr\"{o}dinger 
representation, wave functionals are mappings from a {\it field configuration
space} to the complex plane. The elements of this space, for the time being 
denoted as $\varphi$, are functions from the space manifold into some 
target manifold. In general, there is a Hamiltonian
\beq
H=e_{0} T+\frac{1}{e_{0}} V[\varphi] \;, \nonumber 
\eeq
where the kinetic term $T$ is an operator which does not 
commute with $\varphi$
and 
$e_{0}$ is what we will call the bare 
coupling constant. In fact, $e_{0}$ is not the standard coupling of some field theories, but
we present the Hamiltonian in this form anyway (since it will be convenient in
our analysis of the sigma model). Some regularization is always
present in our discussion.

To be clear as to what we mean by potential energy, the 
potential-energy
operator $V[\varphi]$ contains gradient terms. For example, a 
scalar theory has the potential energy
\beq
V[\varphi]=\int d^{D} x\; \{ \frac{1}{2}[{\bf \nabla}\varphi_{b}(x)]^{2} + {\cal V}(\varphi_{b}(x)) \}\;, \nonumber
\eeq
in dimension $D+1$, where 
the species are labeled by $b$ and ${\cal V}$ is a local potential function.

There is a natural notion of a {\it distance} between any pair of configurations. There is
nothing vague about this idea as we will attempt to show. Suppose the spectrum is 
to be calculated with a strong-coupling expansion (say on a space lattice with continuous time). Such
an
approximation proceeds by first diagonalizing $T$, then treating $V$ as a perturbation. One way
to proceed towards solving for the spectrum is through the representation of
the evolution operator through the Lie-Trotter product formula:
\beq
{\cal U}(t)=e^{iHt}= 
\lim_{N\rightarrow \infty} \left( e^{ie_{o}T\epsilon} e^{iV\epsilon/e_{0}}\right)^{N} \vert_{N\epsilon=t} \;,
\nonumber 
\eeq
which is then expanded in powers of $V$. Such an approximation is very crude
and requires resummation techniques to be made of practical value; but 
such details, though mathematically detailed, are not conceptually sophisticated. In any
case, at leading order, one expects rather generally that the matrix element of the exponential
of the kinetic term between two states
localized in configuration space, is of the form
\beq
<\varphi_{1}\vert e^{ie_{0}T\epsilon}  \vert \varphi_{2}> \sim ({\frac{e_{0}}{2\pi i\,\epsilon}})^{{\cal N}} 
\exp\frac{ie_{0}}{2\epsilon} \rho[\varphi_{1}, \varphi_{2}]^{2}\;, \label{kernel}
\eeq
where 
${\cal N}$ is the (ultraviolet regulated) number of degrees
of freedom. The function of two configurations 
$\rho[\varphi_{1}$, $\varphi_{2}]$ can be identified as a distance. The expression
\rf{kernel} is difficult to evaluate exactly, and it is reasonable to expect that substitution of 
a
function 
$\rho^{\r}[\varphi_{1},\varphi_{2}]$ in place
of $\rho[\varphi_{1},\varphi_{2}]$ 
will make no difference as $\epsilon \rightarrow 0$, provided
that this choice satisfies the criterion that as 
$\rho[\varphi_{1},\varphi_{2}]\rightarrow 0$
\beq
\frac{\rho^{\r}[\varphi_{1},\varphi_{2}]-\rho[\varphi_{1},\varphi_{2}]}
{\rho[\varphi_{1},\varphi_{2}]}
 \rightarrow 0\;.\nonumber
\eeq
In other words, as either distance vanishes, the two distances become
indistinguishable. This indistinguishability has a geometric interpretation. It 
guarantees uniqueness of the infinitesimal metric, which on
physical grounds should have the Riemannian form:
\beq
d\rho^{2} \equiv \rho[\varphi, \varphi+\delta \varphi]^{2}
=\int d^{D}x\, \int d^{D} y\;G_{(x,A)\,(y,B)}\; \delta \varphi^{A}(x)\,\delta \varphi^{B}(y)\;, \nonumber
\eeq
where the symbols $A$ and $B$ include species, Lorentz indices, etc. However, we will show
that the 
general distance function $\rho[\cdot,
\cdot]$ has its uses.

Finding the distance is easy for the example of a 
scalar field
theory. The kinetic term is
\beq
T=-\frac{1}{2}\int d^{D} x \;\frac{\delta^{2}}{
\delta \varphi_{b}(x) 
\delta \varphi_{b}(x)^{*} 
}\;. \nonumber
\eeq
A distance which satisfies \rf{kernel} is given by the Pythagorean
expression
\beq
{\rho[\varphi_{1},\varphi_{2}]}^{2}=\frac{1}{2} \int d^{D} x \;\vert \varphi_{b\;1}(x)-\varphi_{b\;2}(x)\vert^{2}\;. \label{scalar}
\eeq

For a gauge theory in temporal gauge, the physical degrees of freedom are not connections $A_{i}(x)$ (where
$i=1,\dots D$ is a space-coordinate index) in some Lie
algebra, but {\it orbits}. These
are equivalence classes of connections related by gauge transformations. The set whose
elements are $A^{g}_{i}=g^{-1}A_{i}g+ig^{-1}{\partial}_{i}g$ is an orbit $\a$. A choice
of 
distance between two orbits $\a$, $\b$ is 
\beq
\rho[\a,\b]^{2}= \inf_{A\in \a\; B\in \b} \frac{1}{2} \int d^{D} x 
\;tr\,[A_{i}(x)-B_{i}(x)]^{2}\;. \label{yangmills}
\eeq
This expression was considered by Babelon and Viallet \cite {babelon2} and 
later (and somewhat
implicitly)
by Feynman \cite{feynman}. It
was proven to be a metric in the continuum (provided care is taken as to the 
mathematical details) in reference
\cite{orland} and a lattice analogue was discussed in reference \cite{kmo}. Since \rf{yangmills}
depends only on the gauge orbits $\a$ and $\b$ and not on the specific
connections in those orbits, it is gauge invariant. Given any 
metric space, there is a second metric function which can be defined, called
the {\it intrinsic metric} \cite{alex}. This latter metric is 
the length of the shortest path between two configurations (a 
consequence of the triangle
inequality is that the intrinsic metric is always
an upper bound on the original metric). For gauge theories, the intrinsic metric 
coincides with the distance \rf{yangmills} \cite{orland}, as 
had been conjectured earlier
by Babelon and Viallet \cite{babelon2}.

Once a distance satisfying \rf{kernel} is found, other issues
can be addressed. The potential energy can be thought of
as a ``height function" on the configuration space. We would
like to use our intuition about topography, {\it e.g.} the
hills, saddles and valleys of the height function on configuration space, to
understand the properties of the spectrum of the
quantum theory. We give some examples of the application
of this intuition below.

Our first example is a one-component scalar
field with a convex potential
${\cal V(\varphi)}$. In such a situation, the extension of the
ground-state wave functional from
the origin of configuration space ($\varphi=0$) is controlled
by the potential. Even if the (regularized) theory has a vanishing bare mass 
a gap will appear in the spectrum. Whether the interaction terms in the potential can survive
as the cut-off
is removed (and whether the renormalized theory has a gap) depends upon the
anomalous dimensions of these terms.

For our second example, we shall give a highly nonrigorous proof of
Goldstone's theorem. Suppose, for a multi-component scalar theory
there are inequivalent, continuously parametrizable, degenerate minima of
the potential, characterized by an expectation value $v_{b}=\langle\varphi_{b}\rangle \neq 0$. When
the 
volume is finite, the true vacuum is nondegenerate. Let us consider
the distance between two constant low-potential-energy configurations 
\beq
\varphi_{b\;1}(x)=v_{b}\;, \;\;\varphi_{b\;2}(x)=v_{b}^{\r}\;,\;\; 
\vert v \vert^{2}= 
\vert v^{\r}\vert^{2}\;.\nonumber
\eeq
Then \rf{scalar} becomes
\beq
{\rho[\varphi_{1},\varphi_{2}]}^{2}=\frac{V}{2} \vert v_{b}-v_{b}^{\r}\vert ^{2}\le KV\;. \nonumber
\eeq
where $V$ is the volume of space and $K$ is a constant. The upper bound on quantity diverges in the thermodynamic
limit. In quantum mechanics, a vanishing potential energy on a domain of diameter
$\sqrt{KV}$ implies a gap between the (nondegenerate) ground state and the first excited state of order
$1/\sqrt{V}$. This gap vanishes and, by Lorentz invariance, the 
spectrum is completely continuous as $V \rightarrow \infty$.

Actually the argument of the previous paragraph is a bit too crude. There are situations
in which breaking of a continuous symmetry does not occur, despite
the fact that the distance between pairs of low-potential-energy configurations is unbounded in
the thermodynamic limit. In
particular, when
$D=1$, a state localized near a given value of $\varphi$ has low potential
energy, but quantum fluctuations raise the expectation value of the kinetic
energy. This latter energy is lowered through 
the formation of domain walls (these are point-like objects in one space dimension) between regions
of different values of $v_{b}$. These domain walls cost almost no potential energy, as will be shown
explicitly for the one-dimensional sigma models in sections 7 and 12. Their potential
energy turns out to be of order $1/L$, where $L$ is the one-dimensional
volume. Since they also lower the expectation value
of the kinetic energy, they condense in the vacuum. We caution the reader that we use the
term ``domain wall" somewhat loosely and only for lack of a better name (more
accurate
but more cumbersome is ``low-energy nonlinear wave"). What we
are calling domain walls are not topologically stable objects, as are the domain walls 
of the Ising model, though we will show that they can be thought of as solitons. When the
space dimension
is two, a domain wall is a one-dimensional object in space. It
now has a potential energy proportional to its length $l$, as well as inversely 
proportional to the diameter of two-dimensional space $L$ (which in turn is proportional to the
square root of the volume). This energy is therefore
roughly $\sim l/L$. This
energy will vanish in the infinite-volume limit 
only if the typical length $l$ of a domain wall is not proportional to $L$. If it does
not vanish, spontaneous
breaking of a continuous symmetry is possible.

We will now summarize the situation for our final examples, the $O(2)$ and $O(3)$ sigma 
models, for the benefit
of the reader who would prefer to wade through the 
physical ideas before plunging into the mathematics. We study the problem of 
minimization of the potential energy for fixed distance from a constant configuration
(the classical vacuum). More precisely, we consider this problem using two different distances, one
of which is the natural distance in the sense we have already 
discussed, as well as a second distance (due to Feynman
\cite{feynman}) which 
is insensitive to global $O(n)$ transformations. For
either distance we find that the potential-energy surface has grooves or {\it river valleys}. The
type of configuration on a river valley is specified by a parameter, or set
of parameters (which are the invariants of an elliptic function \cite{WW}). We now
summarize how the configurations along a river valley appear as these parameters are
tuned. The
configuration begins as a constant or near constant. Then a weak, long-wavelength 
disturbance, or spin wave appears; this is nearly sinusoidal. Next the amplitude
of the spin wave increases, but has nearly vanishing derivatives, except in
a small region. This configuration is the domain wall mentioned in the previous
paragraph. As the configuration continues to move along the river valley, the
domain wall narrows to a region of the size of the short-distance cut-off. Outside
the domain wall, the configuration is nearly constant. At this stage the potential energy 
is enormous, and is only finite by virtue of the cut-off. Remarkably, the distance
between this configuration and the constant configuration is infinitesimal, although
the two configurations are separated by a high, thin potential-energy barrier. At the top
of the barrier, the configuration is not a domain wall at all (the value
of the field on each side of this object is nearly the same!). We will call it the
{\it barrier configuration} (this is probably similar to what Asorey and Falceto
call a ``sphaeleron" in the $O(3)$ model \cite{asorey}).

As we have already stated, the domain walls have nearly zero potential energy and
always condense in the vacuum. Whether or not barrier configurations 
also condense 
depends on the dynamics
of the particular sigma model. For the $O(2)$ model we compute the 
WKB tunneling amplitude through the barrier. We find that
for a sufficiently large bare coupling constant, the barriers do condense. We 
show that
tunneling through a barrier is a vortex \cite{kt}. In this way, we
obtain the first analytic
demonstration
of the phase transition within the Hamiltonian framework of this model. The
$O(3)$ model is harder to deal with. However, we are able to find the explicit
form of the river valleys. They sit together in a higher-dimensional subset
of configuration space, which we call a {\it river delta}. Surprisingly, as 
the potential energy begins to grow, the river valleys in the
river delta coalesce. At the top of the barrier, the configuration is
similar to that of the $O(2)$ model. It is perhaps adequate to describe
the barrier as ``Abelian". While we do not compute the WKB amplitude of
penetrating the barrier, we do calculate the topological charge
of the barrier-penetration process. It is one half.

\section{The sigma-model metrics}

\setcounter{equation}{0}
\renewcommand{\theequation}{3.\arabic{equation}}

We investigate two different choices of distance for the sigma model. Both
of these are metrics, in that they satisfy the axioms for a metric
space. The first metric is simpler and is easier to justify physically
for the sigma model. We call it the physical
metric. The second, invented by Feynman \cite {feynman}
has the advantage of $O(n)$ global invariance. It is defined on
pairs of points of orbit space, which will be defined below. While the second
metric is no more calculationally powerful than the first and 
is less natural from the point of view of the sigma-model
spectrum, we
consider
it for several reasons. First, Feynman claimed that a certain
property is true of the potential-energy surface of the $O(3)$ sigma
model using this
metric. In fact, this is not correct and we feel it is necessary to
explain some of the subtleties as to why not (this is 
done in section 11). Second and more important, Feynman's 
metric is nice from a mathematical
point of view, and it is easier to 
visualize the potential
energy on orbit space than on the space of field configurations.

In this section and the two sections which follow we discuss some general properties
of the metrics 
of the nonlinear sigma model in $D$ space and one time dimension. We specify
the space dimension $D$ to be one in section 6.

Let $s(t,x)$ be a unit real $n$-vector-valued field on some $D$-dimensional
space manifold (not the space-time manifold, whose dimension is $D+1$), written
as 
\beq
s(t,x)= \left( \begin{array}{c} s_{1}(t,x) \\
                                  .      \\
                                  .      \\
                                  .      \\
                                s_{n}(t,x)  \end{array} \right) \;,   \nonumber
\eeq
with $s(t,x)^{T}s(t,x)=1$.

The action of the $O(n)$ sigma model 
is
\beq
S=\frac{1}{2e_{0}}\int dt\,d^{D}x\, 
(\partial_{t} s^{T}\partial_{t} s
-{\bf \nabla}_{x} s^{T}\cdot{\bf \nabla}_{x} s
)\;. \label{action}
\eeq
Wherever a regularization is not explicit in our discussion, it
will usually be implicit.

The Hamiltonian is built from a similar unit-vector $s(x)$, which is a
$c$-number operator, as well as the angular-momentum operator
\beq
L^{k_{1}}(x)=-i \epsilon^{k_{1}\;k_{2}\; \cdot \cdot \cdot\; k_{n} }
s_{k_{2}}(x)\cdot \cdot \cdot s_{k_{n-1}}(x)
\frac{\delta}{\delta s_{k_{n}}(x)}   \;. \nonumber
\eeq
Explicitly, the Hamiltonian is 
\beq
H=\int d^{D}x [\; \frac{e_{0}}{2}L^{T}(x)L(x)+\frac{1}{2e_{0}}
{\bf \nabla}_{x} s(x)^{T}\cdot{\bf \nabla}_{x} s(x)
\;] = e_{0}T
+\frac{1}{e_{0}}U\;.\nonumber
\eeq

The simplest distance one can define is the physical metric
\beq
r[f,s]^{2}=\frac{1}{2}\;\int d^{D} x\;[f(x)-s(x)]^{T}[f(x)-s(x)] \;, \label{physical}
\eeq
this is the integral of the chord between two points on a sphere. It does
not 
actually satisfy the usual properties of a metric space, unless one
is willing to use a certain amount of real analysis. For example, it
is necessary to identify two field configurations which are the same
except on a set of measure zero (this is similar to how 
Hilbert-space vectors are 
constructed from Schr\"{o}dinger wave functions
in
ordinary quantum mechanics). Alternatively, one can define the
lattice metric over lattice points of space $x$:
\beq
r[f,s]^{2}=\frac{a^{D}}{2}\;\sum_{x}\;[f(x)-s(x)]^{T}[f(x)-s(x)] \;, \label{physlatt}
\eeq
where $a$ is the lattice spacing. In any case, with the appropriate
definition of \rf{physical} or with \rf{physlatt} it is easy to
show the three metric properties
for field configurations $s$, $f$ and $g$: reflexivity
\beq
r[f,s]=r[s,f] \;, \nonumber
\eeq
positivity
\beq
r[s,f]\ge 0 \;, \label{positivity}
\eeq
with equality holding only if $s=f$, and the
triangle inequality
\beq
r[s,f]+r[f,g] \ge r[s,g]\;. \label{triangle}
\eeq

Just as for gauge theories, one can define an orbit space
for the sigma model. This is defined as a set of
equivalence classes of field configurations. Two
field configurations $s$ and $f$ are said to be {\it equivalent} if $f(x)
=R\;s(x)$, where $R\in O(n)$ is a global rotation. This relation is obviously an equivalence
relation. Each equivalence class is an {\it orbit}. If the set of unit vector fields is
called $\cal S$, the {\it orbit space} ${\cal M}={\cal S}/O(n)$ is
the set of orbits
\beq
\psi=\{\;Rs(x):R\in O(n)\; \} \;.   \nonumber
\eeq
Two sigma-model fields
$s$ and $f$ are equivalent if and only if there exists an orbit
$\psi$ such that $s\in \psi$ and $f\in \psi$.

The use of orbit space
takes some justification. In the sigma model, unlike
in gauge theories, not all states are singlets under the symmetry
group. These states transform under some representation of this group (though
if spontaneous symmetry breaking is absent, the ground state is
a singlet, transforming under the trivial representation). There
is however
an obvious reason for studying the space $\cal M$. The potential energy
is invariant under $O(n)$ transformations. Thus as far as the spectrum
of the potential-energy operator $U$ is
concerned, there is no physical difference
between different elements of $\psi$.

To better motivate the definition of $\cal M$, we define a modified
sigma model, which has a real antisymmetric $n\times n$-matrix
gauge field ${\cal A}(t)$ depending
on time, but not on space. The Euclidean path integral is
\beq
Z&=&
    \int {\cal D} s(x,t)\;\delta(s^{T}s-1)\; \int {\cal D} {\cal A}(t)
    \exp-\int d^{D}x \int dt \{ \frac{1}{2e_{0}}
    \vert [\partial_{t}-{\cal A}(t)] s(x,t) \vert^{2}   \nonumber \\
&+&\nabla s(x,t)^{2}  \} \;. \label{2.1}
\eeq
The gauge field has only a finite number (namely
$n(n-1)/2$) of degrees
of freedom. The path integral (\ref{2.1}) is not Lorentz
invariant. Nonetheless, the Green's functions are the
same as those of the usual sigma model in the thermodynamic
limit. For, by making a suitable gauge transformation
${\cal A}(t) \rightarrow q^{-1}(t) {\cal A}(t) q(t)-iq^{-1}(t)q(t)$ and
$s(x,t)\rightarrow q^{-1}(t)s(x,t)$, where
$q(t)$ is an orthogonal matrix, ${\cal A}$ can be set to zero
for almost every time $t$. Thus (\ref{2.1}) can be reduced to the usual
path integral except at a boundary chosen at some particular value of $t$.

There is a lattice version of (\ref{2.1}). The continuum
coordinates $x,t$ will now be replaced by lattice coordinates. These
will also be written as $x,t$, but are integers, equal to the
corresponding
continuum coordinates divided by the lattice spacing $a$. The
lattice path integral is
\beq
Z
=\left[ \prod_{x,t} \int d^{n}s(x,t)\;\;\delta(s(x,t)^{T}s(x,t)-1)\, \right] \nonumber 
\eeq
\beq
&\times& \left[\prod_{t} \int dR(t) \right] \exp-\frac{a^{D-1}}{2e_{0}}\sum_{x,t}
        \{ [s(x,t+1)-R(t)s(x,t)]^{T}[s(x,t+1)-R(t)s(x,t)]    \nonumber \\
&+     &\sum_{i=1}^{D}[s(x+i,t)-s(x,t)]^{T}[s(x+i,t)-s(x,t)]\}\;, \label{2.2}
\eeq
where the measure of integration over the lattice gauge field
$R$ is the $O(n)$ Haar measure. Let us ask
how the variable $R(t)$ behaves if no gauge fixing is imposed. It
is clear that in the semiclassical approximation, the configurations
with $R(t)$ chosen to minimize
$\sum_{x}[s(x,t+1)-R(t)s(x,t)]^{T}[s(x,t+1)-R(t)s(x,t)]$ will
dominate. Suppose then that for some fixed choice of $t$ and
$s(x,t)$, $R(t)={\bar R(t)}$ is this minimum. Note that ${\bar R}$ is 
unique. This is because in the
continuous-time limit ${\bar R}$  is the solution of a first-order
ordinary
differential equation. The 
fluctuations around ${\bar R}$ have a coefficient proportional
to the size of physical space. Consequently, these 
fluctuations are 
suppressed in the thermodynamic 
limit. Therefore (\ref{2.2}) can be replaced 
by
\beq
Z
&=     &\left[\prod_{x,t} \int d^{n}s(x,t)\;\;\delta(s(x,t)^{T}s(x,t)-1)\,\right]     \nonumber \\
&\times&        \exp-\frac{a^{D-1}}{2e_{0}}\sum_{x,t}
        \{ [s(x,t+1)-{\bar R(t)}s(x,t)]^{T}[s(x,t+1)-{\bar R(t)}s(x,t)]    \nonumber \\
&+     &\sum_{i}[s(x+i,t)-s(x,t)]^{T}[s(x+i,t)-s(x,t)]\}\;, \nonumber
\eeq

Thus the path integral may be regarded as the Wick rotation of quantum mechanics of
a particle with mass $e_{0}^{-1}$
in a space in which 
the distance is $\rho[\cdot, \cdot]$, defined by
\beq
\rho[\phi, \psi]^{2}=\frac{a^{D}}{2}\inf_{R\in O(n)}\;\sum_{x}\;
[Rf(x)-s(x)]^{T}[Rf(x)-s(x)] \;, \label{2.10}
\eeq
where $f(x)\in \phi$ and $s(x)\in \psi$. The classical kinetic energy of a 
time-dependent orbit is the
arc-length of the curve this orbit ``traces out" in orbit space.

We will now 
show that on the lattice $\rho[\cdot,\cdot]$ is a metric on orbit space. The 
orbit space is therefore a metric space.  There are three properties
which must hold for this to be true. Obviously for any two 
orbits $\phi$ and $\psi$,
\beq
\rho[\phi, \psi]=\rho[\psi, \phi] \;. \label{refl}
\eeq
It is straightforward to check that for these orbits
\beq
\rho[\phi, \psi]\ge 0 \;, \label{pos}
\eeq
with equality holding only if $\phi=\psi$. Finally, suppose that $\phi$, $\psi$ and
$\tau$ are orbits. Let 
$f(x)$, $s(x)$ and $t(x)$ be unit vector fields
on the
lattice such 
that $f(x)\in \phi$, $s(x)\in \psi$ and $g(x)\in \gamma$. The triangle
inequality \rf{triangle} for field configurations
implies
\beq
\sum_{x}\;
[Rf(x)-s(x)]^{T}[Rf(x)-s(x)]
&+&  \sum_{x}\;
     [s(x)-R^{\prime}g(x)]^{T}[s(x)-R^{\prime}g(x)]   \nonumber \\
&\ge&\sum_{x}\;
[Rf(x)-R^{\prime}g(x)]^{T}[Rf(x)-R^{\prime}g(x)]\;,  \nonumber
\eeq
for any $R$, $R^{\prime}$ in $O(n)$. Taking the greatest lower bound of the sum of first
two terms over $R$ and $R^{\prime}$ gives
\beq
\rho[\phi, \psi]+ \rho[\psi, \gamma] \ge \rho[\phi, \gamma]\;. \label{tri}
\eeq
Now that the three properties (\ref{refl}), (\ref{pos}) and (\ref{tri}) are proved, it
follows that
$\rho[\cdot,\cdot]$ is a metric and that orbit space is a metric space.

The continuum
limit of (\ref{2.10}), dropping factors of the lattice spacing, is
\beq
\rho[\phi, \psi]^{2}=\inf_{R\in O(n)}\;\frac{1}{2} \int d^{D} x\;
[Rf(x)-s(x)]^{T}[Rf(x)-s(x)] \;. \label{2.11}
\eeq
This expression coincides
with the ``minimal distance" for the sigma
model 
Feynman attempts to estimate in the appendix to his 
article \cite{feynman}.

Strictly speaking, in the
continuum neither \rf{physical} nor
(\ref{2.11}) is actually a metric, unless we are careful about configurations
differing on sets of measure zero (otherwise, the inequalities \rf{positivity} and
\rf{pos} can be saturated by
distinct configurations). If this issue is properly dealt
with, the potential-energy function is infinitely discontinuous on the space
of configurations (this will be explained later in this article). However, a 
regularization
will make this discontinuity finite or can remove it altogether.

\section{Evaluation of Feynman's orbit space metric}

\setcounter{equation}{0}
\renewcommand{\theequation}{4.\arabic{equation}}

Unlike the case of the Yang-Mills metric, it is possible to evaluate
\rf{physical} (and (\ref{2.10})) explicitly. Let us first rewrite \rf{physical} as
\beq
\rho[\phi, \psi]^{2}=\inf_{R\in O(n)}\; \int d^{D} x\;
[1-s(x)^{T}Rf(x)]\;.             \label{3.1}
\eeq
Define the matrix $M$ by
\beq
M_{kl}=\int d^{D}x f_{k}(x)s_{l}(x)  \;.    \nonumber
\eeq
Then
\beq
\rho[\phi, \psi]^{2}=V-\sup_{R\in O(n)}\;tr\,RM\;,          \nonumber
\eeq
where $V$ is the volume of space. The $O(n)$-covariant variation
of $R$ is $R\rightarrow \delta \g R$, where $\delta \g$ is an infinitesimal
antisymmetric matrix. The condition that $tr\,RM$ be a local extremum
is therefore that $RM$ is a real
symmetric matrix, {\it i.e.} $(RM)^{T}=RM$. Therefore, there exists 
an orthogonal matrix $P$ which diagonalizes $RM$:
\beq
RM=P^{T} \Lambda P\;,\;\; \Lambda
=
\left( \begin{array}{cccccc}
\lambda_{1} & 0           & \cdot & \cdot    & \cdot & 0 \\
0           & \lambda_{2} & \cdot & \cdot    & \cdot & 0 \\
\cdot       & \cdot       & \cdot & \;\;     & \;\;  & \cdot \\
\cdot       & \cdot       & \;\;  & \cdot    & \;\;  & \cdot  \\
\cdot       & \cdot       & \;\:  & \;\;     & \cdot & \cdot     \\
0           & 0           & \cdot & \cdot    & \cdot & \lambda_{n}
                                \end{array} \right) \;.   \nonumber
\eeq
Then $tr\,RM=\sum_{k} \lambda_{k}$. In
fact, the absolute maximum of the trace coincides with the
supremum. For this case, all the diagonal entries in $\Lambda$
are nonnegative. To see that this is so, assume the contrary. If
$\lambda_{k}$ is
negative for a particular $k$ between one and $n$, $R$ can be replaced by
another $O(n)$ matrix:
\beq
R\rightarrow
\left( \begin{array}{cccccccc}
1 & \; & \;    & \;   & \;& \;    & \;    & \;          \\
\;& 1  & \;    & \;   & \;& \;    & \;    & \;        \\
\;& \; & \cdot & \;   & \;& \;    & \;    & \;        \\
\;& \; & \;    &\cdot & \;& \;    & \;    & \;       \\
\;& \; & \;    & \;   &-1 & \;    & \;    & \;        \\
\;& \; & \;    & \;   & \;& \cdot & \;    & \;        \\
\;& \; & \;    & \;   & \;& \;    & \cdot & \;        \\
\;& \; & \;    & \;   & \;& \;    & \;    & 1
                                \end{array} \right) \;R\;,   \nonumber
\eeq
where the $-1$ appears in the $k^{th}$ row and $k^{th}$ column. Such a
transformation must increase the value of $tr\,RM$. Thus, if $R$ is
chosen to
absolutely maximize the number $tr\,RM$, all the entries of $\Lambda$
are positive definite. Now $M^{T}M=P^{T}\Lambda^{2}P$, and
the matrix $RM$ has only positive
eigenvalues. Therefore
\beq
tr\,RM= tr\sqrt{M^{T}M}  \;. \label{3.2}
\eeq
The meaning of (\ref{3.2}) should be clear; the right-hand side is the
sum of the square roots of the eigenvalues of the symmetric matrix
$M^{T}M$. Since
this expression is unique, it must coincide with the supremum. Thus
(\ref{3.1}) reduces to
\beq
\rho[\phi, \psi]^{2}=V-tr\,\sqrt{M^{T}M}=V-tr\, \sqrt{\int d^{D}x \int
d^{D}y \;
f(x)[s(x)^{T}s(y)]f(y)^{T} }   \;. \label{3.3}
\eeq

Having carefully derived \rf{3.3}, let us note that in retrospect
it is obvious. It is the only positive-definite
$O(n)$-invariant expression
with the
correct dimensions satisfying $\rho[\psi, \psi]=0$.

There is no essential complication in taking the analysis above to the
lattice. Then (\ref{3.3}) is replaced by
\beq
\rho[\phi, \psi]^{2}={\cal L}-tr \sqrt{\sum_{x} \sum_{y} \;
f(x)[s(x)^{T}s(y)]f(y)^{T} }   \;, \nonumber
\eeq
where $\cal L$ is the total number of lattice sites. This expression is equal to
(\ref{2.10}).

\section{The infinitesimal metrics} 

\setcounter{equation}{0}
\renewcommand{\theequation}{5.\arabic{equation}}

In this section, we discuss the Riemannian form of our metrics for
small separations. To do this rigorously, a
lattice or alternatively Hilbert-space techniques must be used
\cite{orland}. We leave it to the more mathematically oriented reader to
fill in the gaps. This section is not necessary 
to understand the rest of the article. However, in the
arguments and calculations which follow we
often discuss curves and functions on configuration space. While these
can be made well-defined concepts on general metric spaces \cite{alex}, the
reader may feel more comfortable knowing that differential-geometric
intuition applies as well.

The functional Riemannian metric on configuration space is
directly written down from \rf{physical}
\beq
dr^{2}=\frac{1}{2} \int d^{D} x \;\delta s^{T}(x) \delta s(x) \nonumber
\eeq
where $\delta s$ satisfies
\beq
\int d^{D} x\; \delta s^{T}(x) s(x) =0 \nonumber\;.
\eeq
There is no more work to be done for this case.

The
functional Riemannian metric on orbit space has the unusual feature
that the
metric
tensor has zero eigenvalues. The corresponding zero modes are the
directions along symmetry transformations. In other words, these
zero modes are motions
along the fibers of the fiber bundle of field configurations. The
reader can find a general discussion of such metric tensors in the
appendix of reference \cite{orland}, though the essential idea
of using a bilinear form with zero eigenvalues as an inner product
was discussed much
earlier 
\cite{singer}. Since the sigma-model 
orbit space is not very difficult
to understand, the standard of rigor in this section is
considerably less than that used for the Yang-Mills orbit space \cite{orland}.

Let us find the infinitesimal form of the metric on orbits $\phi$
and $\psi$ which are infinitesimally close, {\it i.e.}, (\ref{3.3}) is very
small. Then there must exist representatives of $\phi$
and $\psi$, which we call $f$ and $s$ and which satisfy 
\beq
s=f+\delta f\;.   \nonumber
\eeq 
Given $f$, we can in principle find $\delta f$ by rotating $s$ by an
$O(n)$ transformation minimizing $\int (f-s)^{2}$.

Since $f$ and $s=f+\delta f$ are unit vectors, we can
write $\delta f$ in terms of an antisymmetric tensor $W$ of rank $n-2$:
\beq
\delta f^{i}(x)= \epsilon^{i j_{1} j_{2} \cdot \cdot \cdot j_{n-1}}\;
W_{ j_{1} j_{2} \cdot \cdot \cdot j_{n-2}}(x) \;f^{j_{n-1}}(x)\;. \nonumber
\eeq

Global $O(n)$
transformations can be parametrized as $R=e^{h}$, where $h$ is
a constant antisymmetric matrix, parametrized by a constant tensor $b$
of rank $n-2$ as
\beq
h^{ik}=\epsilon^{i j_{1} j_{2} \cdot \cdot \cdot j_{n-2} k}\;
b_{j_{1} j_{2} \cdot \cdot \cdot j_{n-2}}. \nonumber
\eeq

The metric evaluated on $\phi$ and $\psi$ is 
\beq
d\rho^{2}=\rho[\phi,\psi]^{2}
&=& \min_{b} \frac{1}{2} \int\;d^{D}x
    \;\{ \epsilon^{i j_{1} j_{2} \cdot \cdot \cdot j_{n-2} k}
     [\;W_{ j_{1} j_{2} \cdot \cdot \cdot j_{n-2}}(x) -
     b_{j_{1} j_{2} \cdot \cdot \cdot j_{n-2}}\;]f(x)_{k}\}^{2} \nonumber \\
&=&\min_{b} \frac{1}{2} \int\;d^{D}x
\;[\;W_{ [j_{1} j_{2} \cdot \cdot \cdot j_{n-2}]}(x)-
b^{[j_{1} j_{2} \cdot \cdot \cdot j_{n-2}]}\;]^{2} \;,\label{4.1}
\eeq
where square brackets around indices denote antisymmetrization.

Carrying out the minimization with respect to $b$ in (\ref{4.1}) is straightforward. The
result for the $O(2)$ sigma model is
\beq
d\rho^{2}=\int d^{D}x \int d^{D}y\;G_{\a(x) \a(y)}\; \delta\a(x)\; \delta\a(y)\;, \nonumber
\eeq
where
\beq
G_{\a(x) \a(y)}=\frac{1}{2}\delta^{D}(x-y)- \frac{1}{2V} \;, \nonumber
\eeq
and the angle $\a(x)$ is 
defined by $f_{1}(x)=\sin \a(x)$ and $f_{2}(x)=\cos \a(x)$. Notice that this functional
metric tensor $G$ has no dependence on $\a$, and is therefore flat. The zero modes
are constant rotations of $\a(x)$. For the $O(3)$ model
\beq
d\rho^{2}=\int d^{D}x \int d^{D}y\;G_{ s^{i}(x) s^{j}(y) } \delta f^{i}(x) \delta f^{j}(y)\;, \nonumber
\eeq
where
\beq
G_{ s^{i}(x) s^{j}(y) }=
\frac{1}{2}\delta^{D}(x-y)\delta_{ij}-\frac{1}{2V} \epsilon_{ikl}f^{k}(x) (B^{-1})^{lm} 
\epsilon_{jmr}f^{r}(y)\;, \nonumber
\eeq
where the three-by-three matrix $B$ is
\beq
B^{kl}=\delta^{kl}-\frac{1}{V}\int d^{D}z\; f^{k}(z)\; f^{l}(z)\;. \nonumber
\eeq
It can be checked that the zero modes of the metric tensor are $O(3)$ rotations. The metric
tensor considered as a matrix in function space is idempotent, $G^{2}=G$. A general treatment
of how the Laplacian and the curvature may be determined is given in the appendix
to reference \cite{orland}. One can go further in the case of the $O(3)$ sigma model 
and write the metric tensor in angular coordinates instead of unit-vector coordinates, as
was done above for the $O(2)$ case.

\section{The potential-energy surface and the classical pendulum}

\setcounter{equation}{0}
\renewcommand{\theequation}{6.\arabic{equation}}

Let us denote the
``pure gauge" 
or ``classical vacuum" orbit containing
constant $s_{0}(x)=s_{0}$ by $\psi_{0}$. Consider now the 
following problem for $D=1$ with $V=L$, the length
of one-dimensional space. For
fixed $\rho[\psi, \psi_{0}]$, extremize the potential
energy 
\begin{eqnarray} 
U[\psi]=\int_{0}^{L} \;dx\;\left[ \frac{ds(x)}{dx} \right]^{2},
\;\;s\in\psi\;.\nonumber
\end{eqnarray}  
Let us parametrize $s(x)$ using angles 
$\xi_{1}(x)$,..., $\xi_{n-1}(x)$, by  
\beq
s(x)= \left( \begin{array}{cc} \sin\xi_{1}(x)...\,\sin\xi_{n-1}(x)\\
                               \sin\xi_{1}(x)...\,\cos\xi_{n-1}(x)\\
                                . \\
                                . \\
                                . \\
                               \cos\xi_{1}(x) \end{array} \right)  
\;, \nonumber
\end{eqnarray}  
in the standard way. For $O(2)$ we will label $\xi_{1}$ as $\alpha$ and for $O(3)$ we will label
$\xi_{1}$ and $\xi_{2}$ 
as $\xi$ and $\kappa$, respectively. The problem will 
be considered with the periodic 
boundary condition $s(x)= s(x+L)$, though more general boundary
conditions will also be discussed for the $O(3)$ case 
(in reference \cite{kmo} Neumann boundary conditions were considered).

If $s(x)$ is any representative of $\psi$, then
\beq
\rho[\psi, \psi_{0}]^{2}= L-\vert \int_{0}^{L} s(x)dx \vert\;, \label{sphere}
\eeq
from (\ref{3.3}). The problem is therefore equivalent to the extremization of the 
functional
\beq
A[s]=
\frac{1}{2}\int_{0}^{L} dx\; \frac{ds(x)^{T}}{dx}\frac{ds(x)}{dx}
+\lambda \left[ \;\vert \int_{0}^{L} s(x)dx \vert-v\; \right] \;, \label{5.1}
\eeq
where $v=L-\rho[\psi, \psi_{0}]^{2}$ and $\lambda$ is a Lagrange multiplier. Suppose 
that a solution has been found. Then by rotating this solution, we can make $\int_{0}^{L} s^{j}(x)=0$, for
$j=1$,...,$n-1$, and
$\int_{0}^{L} s^{n}(x)=v$. The solution for the extrema of $U[\psi]$ for fixed $\rho=\rho[\psi,\psi_{0}]$ 
(this defines a sphere whose center is the point $\psi_{0}$) is then
described by the motion of an $n$-dimensional pendulum, which can be written
in terms of elliptic functions and integrals.

We can choose not to work on orbit space but instead asked the
same question on configuration space. In that case we need
to extremize the potential energy, while holding fixed the physical
metric distance from a constant configuration $s_{0}$:
\beq
r^{2}=
\frac{1}{2}\int_{0}^{L}\; [s(x)-s_{0}]^{T}[s(x)-s_{0}] =
-\int_{0}^{L}\; s_{0}^{T}s(x)+L 
\;.    \nonumber
\eeq
Such a constraint included with a Lagrange multiplier is once again
the $n$-dimensional pendulum equation. We therefore see that the solutions
we will obtain for either problem are the same. A solution of the
problem in orbit space is a set of solutions in configuration space; the
latter solutions
are all the same up to global rotations.

An $n$-dimensional pendulum is a 
massive particle on the sphere $S_{n-1}$ embedded in ${\bf R}^{n}$
under the influence of a constant gravitational force. The coordinates
are $s_{i}$, $i=1,...,n$ with $\sum_{i}s_{i}^{2}=1$. For
simplicity, this force can be taken to be along the $n$-axis. 

The significance of elliptic
functions \cite{WW} in the $n$-dimensional pendulum problem 
can be understood 
physically. There is one variable describing the ``height"
of the particle, namely the $n^{th}$ 
coordinate, $s_{n}$. We will explain in a moment why
the evolution of $s_{n}$ is periodic, {\it i.e.}
$s_{n}(t+T)=s_{n}(t)$, where $T$ is the period. The behavior of other
variables
describing the
pendulum (if $n>2$) is not, in general, periodic. However, the
evolution of these variables must satisfy certain
conservation laws, namely the conservation of $(n-1)(n-2)/2$ of the
$n(n-1)/2$ components of
the angular-momentum tensor. The conserved components are
$l_{ij}=s_{i}p_{j}-
p_{i}s_{j}$ where $i$ and $j$ are less than $n$. The other variables
influence $s_{n}$ only through these conserved components. In other words
the equation of motion of $s_{n}$ can be written so that $l_{ij}$
can be substituted for the other degrees of freedom. Any solution of this
equation is obviously periodic. 

Now imagine
turning the pendulum upside
down; this means reversing the direction of the gravitational 
field. The evolution of $s_{n}$ is again periodic with a new period
$T^{\r}$. A field reversal
is equivalent to the Wick rotation 
$t\rightarrow\, i\,t$. The equation of motion 
contains one term with a second time-derivative. Reversing the direction
of the field is equivalent to changing the sign
of this second time-derivative. This implies that $s_{n}(t)$ in the
original
pendulum (that
is, before
it was turned upside down) has an {\it imaginary period} 
$iT^{\r}$ as well as the real period $T$. If this function
has no singularities in the complex $t$-plane other than poles, it must be
an elliptic
function.

\section{River Valleys of the $O(2)$ model and solitons}

\setcounter{equation}{0}
\renewcommand{\theequation}{7.\arabic{equation}}

Next we shall investigate the potential-energy surface of the
$O(2)$ nonlinear sigma model or classical $XY$ model. More
precisely, we shall examine the pendulum solutions which
are the extrema of \rf{5.1}. Depending upon which
is easier, sometimes
we will discuss the extrema in orbit space and sometimes we
will discuss the
extrema in configuration space. As we pointed out in the previous
section, the latter are contained
in the former.

The planar pendulum displays two types of motion, namely
oscillating and
circulating. The lowest-energy
configuration is that for which the pendulum sits at the nadir for
all time. As the energy is increased, the pendulum makes small harmonic
oscillations 
around the nadir, which can be described by trigonometric
functions. Increasing the energy further leads to an increase
in the period, which 
is an elliptic integral, eventually 
leading to a situation in which the pendulum will spend
most of the time near the zenith, but on occasion will rapidly 
sweep out an angle of nearly $2\pi$. A slight further increase in
energy causes the pendulum to pass through the zenith, leading 
to circulating, rather than oscillating motion. The pendulum
still spends most of its time near the zenith, but the motion is now
consistently clockwise or counter-clockwise. As the energy continues to
increase
the angular frequency of this circulating motion becomes
a constant. Once more the motion can be described by trigonometric
functions.

The mathematical translation of the discussion
in the previous paragraph is the following. For the pendulum
Lagrangian 
\beq
L=\frac{1}{2} {\dot \a}^{2}-\mu(1-\cos\a) \;, \nonumber
\eeq
the classical energy is given by
\beq
E_{pend}=\frac{1}{2} {\dot \a}^{2}+\mu(1-\cos\a) \;. \nonumber
\eeq
There are oscillating solutions to the equations of motion
\beq
\a(t)=2\sin^{-1} k\,sn(k{\sqrt{\mu}} t, k)\;, \nonumber
\eeq
where $sn(u,k)$, sometimes written $sn\,u$, is the Jacobi elliptic-sine function, and
$E_{pend}=2\mu k^{2}$. The modulus $k$ is between zero and one. Increasing the
energy means increasing $k$. The period is $4K$, where $K$ is the complete
elliptic integral 
$K(k)=sn^{-1}1$. This diverges logarithmically as $k$ approaches one (asymptotically,
$K\simeq \log 4/k^{\r}$, where $(k^{\r})^{2}=1-k^{2}$). At $k=1$, these
solutions are joined onto the $k=1$ circulating solutions
\beq
\a(t)=2\sin^{-1} \vert sn(k^{-1}\,{\sqrt{\mu}} \,t, k)\;, \vert \nonumber
\eeq
where $E_{pend}=2\mu/k^{2}$. The reason for the absolute value in this expression is that
there are no turning points and the left-hand side is discontinuous with a discontinuity of $2\pi$. As 
$k$ is then decreased the energy
increases
further, the
angular velocity eventually becoming a constant.

Now let us to turn to the extrema of \rf{5.1}, where time $t$ is replaced by space $x$ (up
to a constant) and
kinetic energy (up to another constant) is now the sigma-model potential-energy density. We 
take periodic boundary conditions. Up to global 
rotations $R$ there are two types of solution
labeled by an integer $N=1$, $2$,$.$$.$$.$ . 
\begin{eqnarray} 
\alpha(x)
=\pm \alpha^{osc}_{N}(x,k,x_{0})= 
\pm 2\sin^{-1} k\, sn \left[ \frac{4NK}{L}(x-x_{0}) \right] \;,  \label{soln1} 
\end{eqnarray}
which 
resemble the oscillating solutions of the pendulum and 
\beq
\alpha(x)
=\pm \alpha^{circ}_{N}(x,k,x_{0})= \pm
2\sin^{-1}\vert sn \left[ \frac{2NK}{L}(x-x_{0})\right] \vert \;,  \label{soln2} 
\eeq  
which resemble the circulating solutions of the pendulum. These
extremal curves in orbit space are nicely parametrized by the 
modulus $k$ as shown in Figure 1. Note: not all of these curves are the river
valleys mentioned in section two.

\begin{figure}[t]
\centerline{\epsfxsize=3.5in \epsfysize=2.67in
\epsfbox{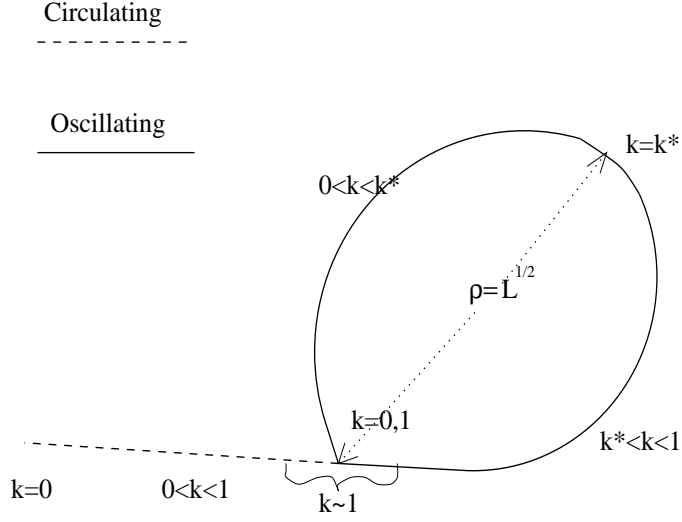}}
\label{1}
\caption[l]{\it Two typical extremal curves for the $O(2)$ sigma model. In the thermodynamic
limit, the 
potential energy is vanishing
along the solid (oscillating) and dashed (circulating) curves where $k$ is not
close to one. The potential energy is ultraviolet divergent in the 
region where $k$ is close to one. }
\end{figure}

We will next investigate the properties of each branch 
of solutions. First let us consider the 
oscillating 
branch (\ref{soln1}). We denote the particular value of $k$ where $E=2K$ by $k^{*}$, where
\beq
E=E(k)=\int_{0}^{1} dn^{2}\,u \,du\;  \nonumber
\eeq
is another 
standard elliptic
integral, not to be confused with the energy
$E_{pend}$. The number $k^{*}$ has the numerical value 
$k^{*}\approx 0.82$. For $0\le k \le k^{*}$
the
distance from $\psi_{0}$ is independent of $N$ and has the form
\beq
\rho^{osc}(k)^{2}=L-\vert \int_{0}^{L} \cos \a^{osc}_{N}dx \vert
=2L\left( 1-\frac{E}{K} \right) \;, \nonumber
\eeq
while for $k^{*} \le k \le 1$ 
\beq
\rho^{osc}(k)^{2}=L-\vert \int_{0}^{L} 
\cos \a^{osc}_{N}dx \vert=2L\frac{E}{K} \;. \label{distance2}
\eeq
The function $\rho(k)$ rises smoothly from $0$ to $L$ as $k$ goes 
from $0$ to $k^{*}$, then
falls off to zero again as $k\rightarrow 1$. Notice that at
$k=1$ the orbit is identical to that at $k=0$. An orbit along a 
extremal curve is maximally
far from the origin at $k=k^{*}$. The potential-energy 
functional $U=\frac{1}{2}\int_{0}^{L} (\frac{d\alpha}{dx})^{2} dx$ for the 
oscillating branch (\ref{soln1}) is
\begin{eqnarray} 
U^{osc}(k)=\frac{32N^{2}K}{L} [E-k^{\prime 2}K] \;, \nonumber
\end{eqnarray}  
where $k^{\prime 2}=1-k^{2}$. For fixed volume $L$, $U^{osc}(k)$ diverges at $k=1$, but, as mentioned 
earlier, this divergence
is regularized by a lattice (or some other ultraviolet cut-off).

Physically the 
oscillating branch of solutions (\ref{soln1}) is a spin wave of 
wavelength $L/N$. As $k\rightarrow 0$, the spin wave dies off in amplitude, approaching $\psi_{0}$. As 
$k\rightarrow 1$, the spin wave begins to {\it wind}. If the angle $\alpha$
is represented on a circle, then the configuration (up to global rotations) is a curve on a cylinder. This is shown 
in Figure 2abc. However the $k=1$ limit is not a simple kink of winding number $N$. For not only is the curve
in Figure 2c beginning to wind, but the amplitude is nearly a constant for all $x$. In other words, the
spin wave has become a domain wall of the type discussed
in 
section 2. The potential energy
of the domain walls, like those of the spin waves is very small. The winding takes place
in narrow regions of space. As $k$ approaches one, the width of these regions collapses to zero. Thus the
$k=1$ solution coincides with the $k=0$ solution. Since the potential energy diverges as $k$ 
approaches one, we can see that
there is an {\it infinite discontinuity} in the potential-energy function on orbit space. This discontinuity
can be removed or made finite with an ultraviolet cut-off.

\begin{figure}[t]
\centerline{\epsfxsize=3.5in \epsfysize=2.67in
\epsfbox{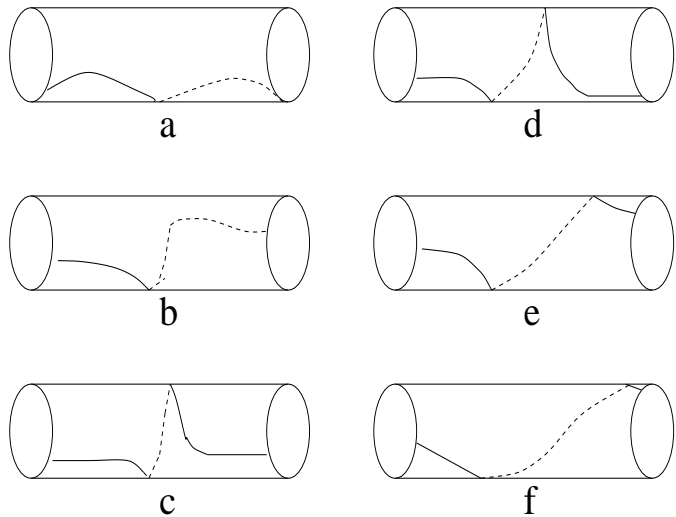}}
\label{2}
\caption[l]{\it How extremal-curve configurations depend on $k$. Here we show a path
in configuration space, starting from $\psi_{0}$, moving away as an
oscillating configuration, reaching $\psi_{0}$ again, then moving away as a circulating
configuration. In each
case, the horizontal axis of the cylinder is $x$, while the angle
is $\a$. For the oscillating solutions, small $k$ implies a small amplitude
spin wave (a), and as $k$ increases the potential energy becomes more
localized into a domain wall (b). When $k$ is close to one, short kinks begin to form (c), while
away from these kinks, the configuration is nearly constant. For the circulating
solutions with $k$ close to one (d), the kinks are fully formed and the winding number
is $N$ 
(the cylinder in this figure is too short to show the remaining kinks). Decreasing
$k$ does not change the number of windings, but the kinks begin to spread out into
domain walls (e). As
$k$ decreases further, the derivatives become of order $1/L$ for all $x$. If
$k\approx 1$ (c and d) the potential energy becomes divergent; otherwise it
vanishes as $L\rightarrow \infty$. }
\end{figure}

For the 
circulating branch the distance from $\psi_{0}$ is given by
\beq
\rho^{circ}(k)^{2}=L-\vert \int_{0}^{L} \cos \a^{circ}_{N}dx \vert
=\frac{2L}{k^{2}}\left( \frac{E}{K}-k^{\r\,2} \right) \;, \nonumber
\eeq
while the potential energy is 
\begin{eqnarray} 
U^{circ}(k)=\frac{8N^{2}KE}{L}  \;. \nonumber
\end{eqnarray}  
Again there is an ultraviolet divergence at $k=1$.

The circulating branch of solutions wind 
around the cylinder (as shown in Figure 2def) $N$ times. In other words, these
are configurations with $N$ kinks. As $k\rightarrow 1$, the potential energy diverges, but the 
orbit approaches $\psi_{0}$. This is because
the regions where the kinks occur will narrow until they disappear. As $k$ decreases, the 
potential
energy decreases as well, but $\rho^{circ}(k)$ becomes of order 
${\sqrt L}$. What happens is that the derivative of $\alpha$ becomes of 
order $\frac{N}{L}$ in this limit.

Not all of the extremal configurations we have constructed
are minima
of the potential energy on a sphere in orbit space. This issue is
examined in detail in Appendix A. We find that all
the circulating solutions
\rf{soln2} are indeed local minima on the orbit-space sphere, for any
value of $k$. Consequently, these are river valleys (note: in 
reference \cite{kmo} the term ``river valley" was used for
all the extremal curves). We also
show that the $N=1$ oscillating solution \rf{soln1} is a local
minimum on the orbit-space sphere for sufficiently small $k$.

We will now interpret our results 
thus far. We have found that the potential energy on 
orbit
space $U[\psi]$ has one-dimensional valleys. In the thermodynamic limit
$L\rightarrow \infty$, the bottoms of these 
valleys are flat (the potential energy vanishes there) except for
the special region where $k$ approaches unity. The
$k=1$ point is actually $\psi_{0}$. The energy of a point in a river valley 
diverges as $\psi_{0}$ is approached. Therefore the potential
energy at $\psi_{0}$ is discontinuous. If a regularization 
is introduced, the energy 
vanishes at $\psi_{0}$, but for $k=1-\epsilon$, where
$\epsilon$ is determined by the 
ultraviolet cut-off, the river-valley potential energy rapidly rises to a large number.
Even in this region, the energy 
must be a local minimum in the direction perpendicular to the river valley.

For large volume, the potential energy is almost 
constant nearly everywhere in a river valley. The one-dimensional
domain where this is so has length $O(\sqrt{L})$. If we 
naively view $k$ as a collective
variable, and ignore fluctuations in other degrees of freedom, the 
gap is of order $O(\frac{1}{L})$.

We note that the extremal curves are not straight lines in 
configuration space. Their 
tangent vectors at $k$ for the oscillating solution \rf{soln1}
are 
\begin{eqnarray} 
\beta^{osc}(x,k)=\frac{\partial \alpha_{1}^{osc}(x,k)}{\partial k}= 
\frac{2sn\,u \;dn\,u-Z(u)cn\,u}{1-k^{2}} \;, \label{tangent}
\end{eqnarray}  
where $u=4K(x-x_{0})$ and $Z(u)$ is the Jacobi zeta 
function. The inner
product of $\beta_{N}$ and its derivative with respect 
to $k$ is not zero, which means that 
the extremal curves have curvature. One can define
the unit tangent vector
${\hat \beta}(x,k)=\beta(x,k)/{\sqrt {{\int_{0}^{L} 
\beta(y,k)^{2}\,dy}}}$.

All extremal-curve configurations, including those of river valleys, are 
related to solitons in a finite
volume. Consider adding to the $O(2)$ sigma model action \rf{action}
an external source $h$ (from the viewpoint of lattice
spin systems, this is an external magnetic field), modifying it to
\beq
S=\frac{1}{2e_{0}}\int dt\,dx\, 
(\,\partial_{t} s^{T}\partial_{t} s
-{\partial}_{x} s^{T}{\partial}_{x} s+ h^{T}s\,
) \;. \label{sine-gordon}
\eeq
This is the sine-Gordon action, with the different classical
vacuua $\alpha(x)=0,\pm 2\pi,\dots$ identified (in this
respect it is closer to the textbook example of the twisting band 
than the
usual sine-Gordon action). Without loss of generality, the direction of $h$ can be chosen 
so that \rf{sine-gordon} reduces to
\beq
S=\frac{1}{2e_{0}}\int dt\,dx\, 
[(\partial_{t} \a)^{2}
-({\partial}_{x} \a)^{2}+h\,\cos\a
] \;, \nonumber
\eeq
and the equation of motion is the sine-Gordon equation, in the
compact field $\a$:
\beq
(\partial_{t}^{2}-\partial_{x}^{2}) \a=-h\sin\a \label{sgeq}\;.
\eeq
Substituting $\a(t,x)=\a(u)$, where
$u=t\pm\frac{x}{v}$ into 
\rf{sgeq} yields 
\beq
\partial_{u}^{2}\a(u)=-\frac{v^{2}h}{(v^{2}-1)} \sin\a \;, \nonumber
\eeq
which is the equation of motion of the planar pendulum. The solutions
to this equation are a train of evenly-spaced solitons, of the
form 
\beq
\a=2\sin^{-1} k\, sn \left( \frac{4Ku}{l}, k \right) \;, \label{soliton1}
\eeq
or
\beq
\a=2\sin^{-1}  \vert sn \left( \frac{2Ku}{l}, k \right) \vert \;, \label{soliton2}
\eeq
where the modulus $k$ is a free parameter between zero and one and
the spacing between solitons is given by
\beq
l={\sqrt \frac{32 K^{2} (v^{2}-1) }{v^{2}h}  } \;. \nonumber 
\eeq
The solitons are located at the nodes $u=0$, $\pm l$, $\pm 2l$,$.$$.$$.$ .
The $k\rightarrow 1$ limit
of either \rf{soliton1} or \rf{soliton2} reduces 
to the standard one-soliton solution. 

Unlike the solutions 
\rf{soliton1} and \rf{soliton2}, the extremal-curve configurations
\rf{soln1} and \rf{soln2} have no explicit time dependence. However, they
do contain the translation parameter $x_{0}$. Thus the soliton solutions
of \rf{sine-gordon} really can be thought of as extremal-curve configurations
with moving $x_{0}$. Conversely, a river-valley configuration is
just a ``snapshot" of a soliton configuration \rf{soliton2} at a given time. The 
solutions \rf{soln1} contain $2N$ solitons (actually they are solitons alternating
with anti-solitons) and \rf{soln2} contain
$N$ solitons.

The reason we emphasize the mathematical
resemblance between extremal-curve configurations and solitons is to give
a physical picture of
quantum barrier penetration, which is discussed in the next section. The
river-valley approaches the trivial configuration $\psi_{0}$ as $k$ tends to
one. Thus a soliton configuration \rf{soliton2}, with diverging potential energy can be made 
arbitrarily close to $\psi_{0}$ (meaning that the metric 
separation in configuration space between the orbit
containing the soliton configuration
and $\psi_{0}$ is small).

\section{Barrier 
penetration and vortices}

\setcounter{equation}{0}
\renewcommand{\theequation}{8.\arabic{equation}}

We have gone to a lot of trouble to investigate the extremal curves, and if the
reader has been patient enough to follow our discussion thus far, we will now
explain their physical significance.

An orbit in an extremal curve has very small potential energy, unless
it approaches $\psi_{0}$ from particular directions, corresponding to $k$ 
close to one. For small $k$ {\it superpositions} of extremal-curve configurations,
\beq
\a=a_{1}\a_{1}+a_{2}\a_{2}+ \cdot \cdot  \cdot \;,\label{super}
\eeq
where the real constants $a_{1}$, $a_{2}$,... are not too large, are also configurations
of small potential energy. There is nothing surprising in this
fact. For small $k$, Jacobi elliptic functions become trigonometric functions
and \rf{super} is a general spin-wave configuration. For small amplitude waves, the
sigma model is approximately a free field theory. However, when $k$ is large, the
extremal-curve configurations cannot be superposed as in \rf{super}, because their nonlinear
character is important.

We now make the following

\noindent {\bf Assertion}: In the Schr{\"o}dinger representation of the $XY$ model, the vacuum wave functional
can be significantly different from zero for only two kinds of
field configurations. These are: \begin{enumerate}
\item Spin waves, {\it i.e.} configurations of the form \rf{super}.
\item River-valley configurations 
with $k$ 
very close to one, as well as slight deformations
of such configurations.
\end{enumerate}
By ``slight deformations" we mean two things:\begin{itemize}
\item First, small-amplitude spin waves can be added to 
large $k$ river-valley configurations without significantly changing the potential
energy. 
\item Second, river-valley configurations are very different from zero only in small
regions of space where kinks begin to appear (we have shown in the last
section that these kinks can be thought of as solitons). There
is no significant gain in either the potential energy or in the distance from $\psi_{0}$ if the 
position of these regions is changed.
\end{itemize}

To see why the assertion must be true, let us first examine each case 1. and 2. individually. Clearly
configurations of type 1. which have small potential energy must exist
as fluctuations around $\psi_{0}$. There is nothing to suppress such fluctuations. Why 
should configurations of type 2. be
important? They should be highly
suppressed
by virtue of their huge potential energy. However, they may not be completely
suppressed. The reason is just 
that the orbits containing these configurations are close to $\psi_{0}$ in orbit space. Even if
quantum fluctuations from $\psi_{0}$ to one of these orbits may be small, there
is the possibility that they do not disappear entirely. Only the configurations 1. and 2. are
either of small potential energy or close to $\psi_{0}$. Therefore the vacuum wave functional
should be vanishly small when evaluated for any configuration other than 1. and 2.

Whether
or not configurations of type 2. indeed appear through quantum fluctuations depends upon 
the barrier-penetration amplitude. It is the computation of this amplitude we
examine in the next section. However, first we will discuss further the physical
interpretation of the barrier penetration.
The energy barriers are more than just the $k\approx 1$ region
of river valleys \rf{soln2}. Consider the point of intersection $\psi$ of a sphere
whose center is $\psi_{0}$ and a river valley. Let {\it G} be a small neighborhood 
of $\psi$ in the sphere (see Figure 3). If this neighborhood {\it G} is sufficiently small, the
point of intersection $\psi$ is actually a {\it minimum} of potential energy in
{\it G}. The $k\approx 1$ part of a river valley is a path of least resistance through the barrier.

\begin{figure}[t]
\centerline{\epsfxsize=4.5in \epsfysize=3.5in
\epsfbox{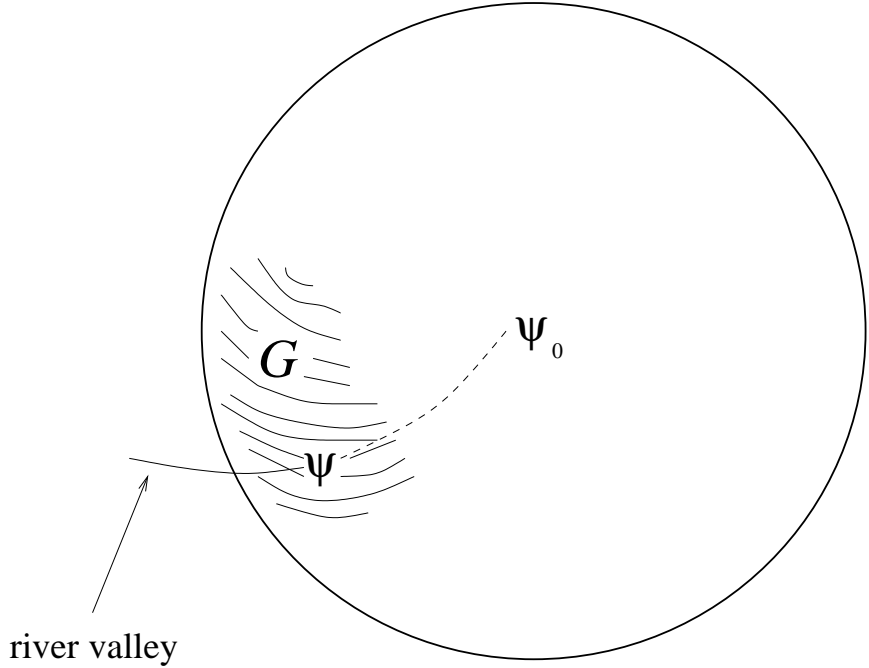}}
\label{3}
\caption[l]{\it River valleys for $k\approx 1$ are 
paths of lowest potential energy through the
ultraviolet divergent energy
barriers. The potential at $\psi$ is a minimum in the neighborhood {\it G}
in the sphere of constant metric distance from $\psi_{0}$.
}
\end{figure}

A typical tunneling event can be visualized with the aid of either Figures 2abc or
2def. A constant configuration makes a transition to the configuration 2c (2d), which
contains short localized regions of large potential energy. These regions
stretch out in physical space, meaning that 
the configuration evolves first to 2b (2e), then 2a (2f) and
finally becomes nearly asymptotically constant (winding with minimal derivatives with
respect to $x$).

The amount of time required to make the initial transition to 2c (2d) depends
on the coupling $e_{0}$ and the nature of the cut-off. In a sensible regularization, the
potential energy in the small region $1>k>1^{*}$ can be set to zero, where $1^{*}$ is a number
very close to
one but strictly less than one. The energy barrier is thereby rendered finite. If the
initial state is a delta function localized in $\psi_{0}$, {\it i.e.} $k=1$, the orbit will
lie in a spreading wave packet, which will eventually 
reach $1^{*}$. Now $e_{0}^{-1}$ plays the role of mass in this quantum-mechanical system. The form
of this wave packet is therefore, for short times,
\beq
\Psi(k,t)= 
{\sqrt{\frac{1}{2\pi i\,t e_{0}^{-1}}}} 
\exp\frac{i\,\rho(k)^{2}}{2t\, e_{0}^{-1}}\;. \label{wavepacket}
\eeq
The separation of these two
points in
orbit space is $\rho(1^{*})$. Therefore, the typical time an orbit takes to travel from $k=1$ to
$k=1^{*}$ is proportional to $e_{0}\rho(1^{*})^{2}$. For small $e_{0}$ the transition is very sudden, and
the tunneling process (shown on a lattice) resembles that of Figure 4. At stronger coupling, this
transition time is large, but still finite.

\begin{figure}[t]
\centerline{\epsfxsize=3.5in \epsfysize=2.67in
\epsfbox{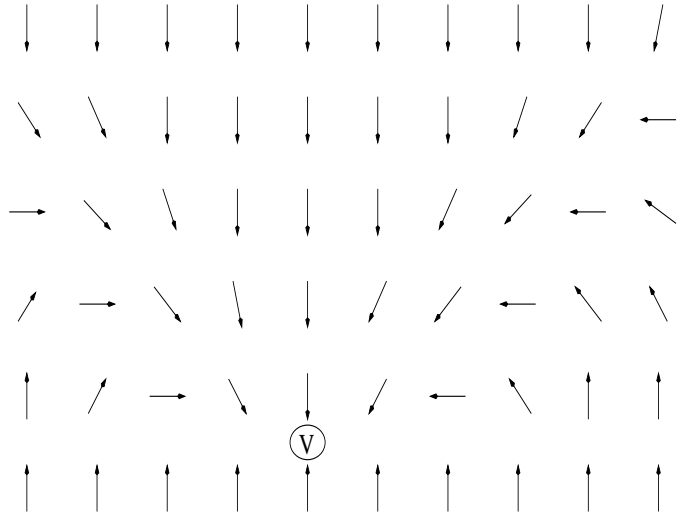}}
\label{4}
\caption[l]{\it The tunneling process at weak coupling. Time is
represented by the vertical direction, and a lattice is used for 
ease of visualization. The initial configuration is constant. The 
vortex (located at the letter V) appears suddenly. Then the 
configuration slowly decays back to a constant. As the
coupling 
increases, the number of lattice spacings between the initial constant
configuration and the vortex also increases.}
\end{figure}

The tunneling process is a {\it vortex} \cite{kt}. The closed line integral of
the gradient of $\a$ enclosing the ``core" of the vortex 
$\int {\bf \nabla} \th \cdot d{\bf l}$, where the topology of the
one-dimensional configuration changes, is $2\pi$. From Figure 4, the reader can see that our
vortex is asymmetrical, unlike Berezinskii-Kosterlitz-Thouless vortices, which
are nearly rotational invariant (Figure 5). In the tunneling process corresponding to the latter, the topology
change does not happen suddenly, even for small $e_{0}$. The two types of vortices do 
not resemble each other very much, except at infinitely-strong coupling. Actually, Euclidean
field configurations
which dominate the lattice partition function
are not isolated vortices, but a superposition of vortices along
with spin waves. Most of these resemble neither our asymmetrical vortex of Figure 4 nor
the more standard variety of Figure 5.

\begin{figure}[t]
\centerline{\epsfxsize=3.5in \epsfysize=2.67in
\epsfbox{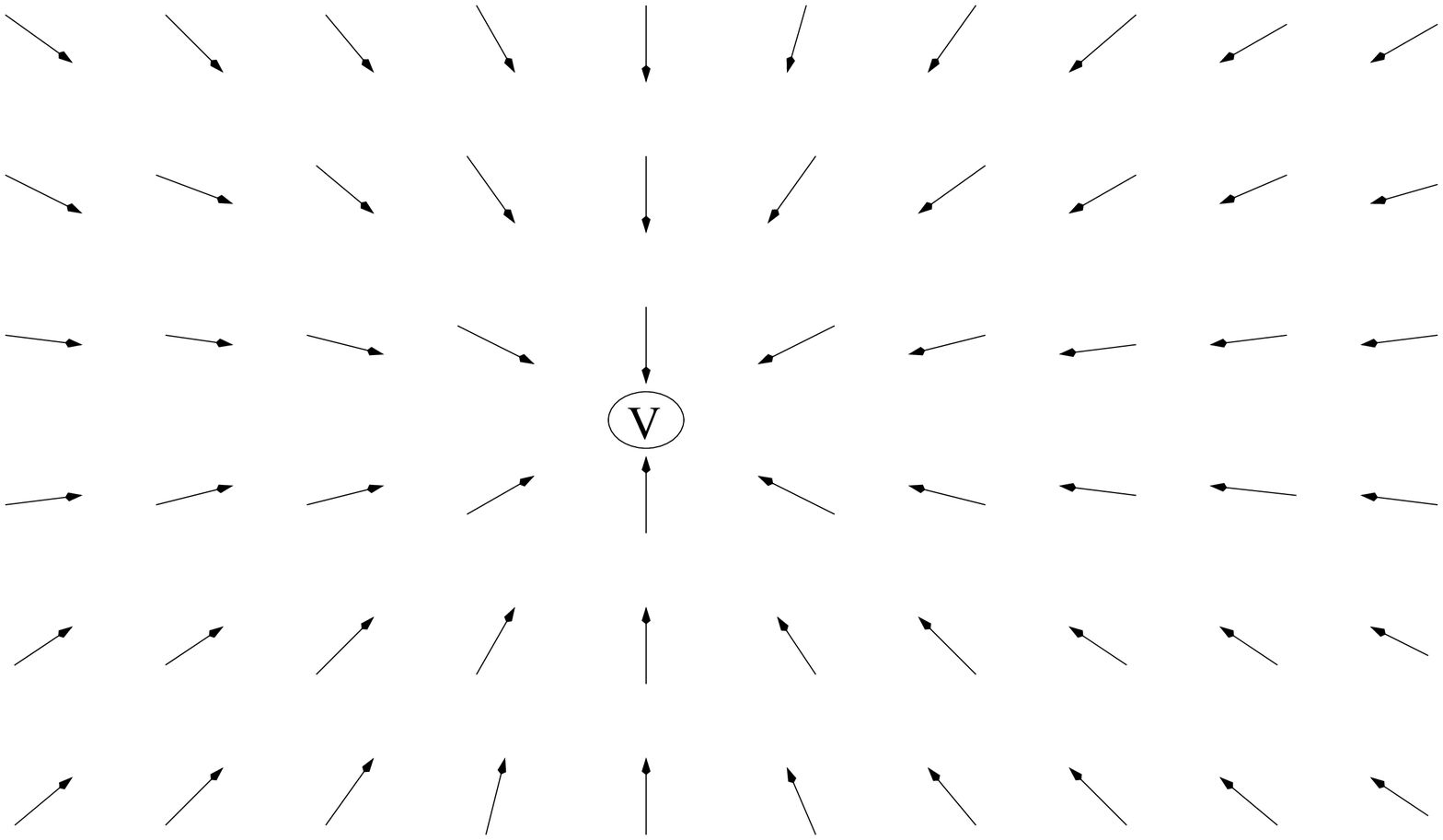}}
\label{5}
\caption[l]{\it The Berezinskii-Kosterlitz-Thouless vortex. The
picture of the tunneling
process is similar, except that it is nearly rotationally
invariant. The vortex appears halfway between constant
configurations during a long time
evolution.}
\end{figure}

\section{The tunneling amplitude in the $O(2)$ model}

\setcounter{equation}{0}
\renewcommand{\theequation}{9.\arabic{equation}}

The ultraviolet divergence in the barrier height is due to the fact that
the derivative 
$\frac{\partial \alpha(x,k)}{\partial x}$ 
diverges as $k \rightarrow 1$. Any sensible regularization
imposes the restriction 
\beq
\vert \frac{\partial \alpha(x,k)}{\partial x}\vert\le \frac{2Q}{a}\;, \label{cut-off}
\eeq
where $a$ is the short-distance cut-off and
$Q$ is a constant depending on the details of the 
regularization. For example, on a lattice, $a$ is
the lattice spacing and the difference
in $\alpha$ at adjacent lattice sites is at most $\pi$, so
that
$Q=\pi/2$.

For the circulating river-valley solution with one soliton, that is
\rf{soln2} with $N=1$, the condition \rf{cut-off} implies that
for each $x$
\beq
\frac{2K}{L}\vert \; dn \left[ \frac{2K}{L}(x-x_{0})\right] \;\vert \le \frac{2Q}{a}\;, \nonumber
\eeq
or
\beq
K(k)\le \frac{QL}{a} \equiv K(1^{*}) \;. \nonumber
\eeq

The path in orbit space of 
the tunneling is along the river valley from $k=1^{*}$
to some value of $k={\tilde k}$ less than $1^{*}$, for fixed $x_{0}$. Recall
that the role of the mass is played by $e_{0}^{-1}$. The
WKB formula for the tunneling amplitude for fixed $x_{0}$ is
\beq
{\cal T}=\exp-W\;, \;\; W=\int_{\tilde k}^{1^{*}} dk\; {\sqrt{\int_{0}^{L}\beta(x,k)^{2}dx}}
\;\;{\sqrt{\frac{2}{e_{0}} \left[
\frac{U(k)}{e_{0}}
-\frac{U({\tilde k})}{e_{0}}\right]  }}\;, \label{WKB}
\eeq
where $\beta$ is the tangent vector to the river valley defined in \rf{tangent}. The square
root of the length of the tangent vector is included in the integrand so that the integration 
measure is
the orbit-space line element.

A useful approximation is the replacement
\beq
dk {\sqrt{\int_{0}^{L}\beta(x,k)^{2}dx}}\; \longrightarrow \; d\rho(k)=\frac{d\rho(k)}{dk}dk \nonumber
\eeq
in \rf{WKB}. This approximation is valid because what remains in the
integrand is small unless $\rho(k)$ is small. In this regime, $\rho(k)$ is
given by \rf{distance2}.

It is convenient to 
use the formulas for the derivatives of the complete elliptic
integrals of the first and second kind 
\beq
dE=\frac{E-K}{2k^{2}} d(k^{2})\;,\;\; 
dK=\frac{E-k^{\prime\;2} K}
{2k^{2}k^{\prime\;2}}
d(k^{2}) \;, \nonumber
\eeq
the approximations
valid when $k\approx 1$ that
\beq
E\approx 1\;,\;\; K\approx \log\frac{4}{k^{\prime}}\;, \nonumber
\eeq 
and to make a change of variable from $k$ to $K$. We also assume that 
in the dominant part of the range of integration $U({\tilde k})$
can be neglected. Then \rf{WKB} becomes 
\beq
W\approx \int_{K({\tilde k})}^{\frac{QL}{a}} \,\frac{2{\sqrt{2}}}{e_{0}K} \,dK \approx\frac{{2\sqrt{2}}}{e_{0}} 
\log\frac{QL}{a}\;.
\label{infrared}
\eeq

Notice that \rf{infrared} implies that the WKB tunneling amplitude at any given
point vanishes in the thermodynamic limit. The
alert reader may have noticed
that $W$ is similar to the usual expression
for the vortex action \cite{kt}.

\section{The phase transition in the $O(2)$ model}

\setcounter{equation}{0}
\renewcommand{\theequation}{10.\arabic{equation}}

We will now explain the Kosterlitz-Thouless
transition using our Hamiltonian quantum-mechanical methods. 

Thus far we have considered a change in the winding around the cylinder at
a specific location. To determine the total transition amplitude it is 
necessary
to sum over all the possibilities of this location. On a lattice, the
number of such possibilities is $L/a$. With a different
regularization, this number should be $L$ divided by the 
short-distance cut-off and multiplied by some (non-universal) constant $Y$. In
any case
\beq
\frac{L}{a} {\cal T}= Y\;Q^{-\frac{ 2{\sqrt{2}} }{e_{0}}} 
\left(\frac{L}{a} \right)^{1-\frac{2{\sqrt{2}}}{e_{0}}}  \;. \label{kt-trans}
\eeq

In the thermodynamic limit 
$L \rightarrow \infty$ this amplitude vanishes unless $e_{0}$ is
greater than its critical value $2{\sqrt{2}}$ 
(as pointed out earlier, this number is
not universal). For small $e_{0}$, tunnelings 
are rare and the vacuum wave functional
is Gaussian. In this weak-coupling 
phase, winding modes or solitons
are stable particles. For sufficiently large $e_{0}$ the solitons condense in
the vacuum, disordering 
correlation functions. The quantity in
the exponential of \rf{kt-trans} is
really
not very different from Kosterlitz and
Thouless' famous mean-field
free-energy estimate. Having said this, we started from
a different point of view, which will later 
prove useful for the 
$O(3)$ sigma model.

Many years ago, soliton condensation in the Schr{\"o}dinger picture
was argued to be responsible for the phase transition of the Hamiltonian
$XY$ model \cite{frad and suss}. Our calculation shows
that this is the case. Our solitons \rf{soliton1}, \rf{soliton2}
are {\it not} classical solutions of the $XY$ model, but the associated model 
\rf{sine-gordon}. We will later argue that soliton condensation takes place
in the $O(3)$ sigma model for any coupling (at topological angle $\th=0$). Once
again, these are not solitons of the classical $O(3)$ sigma model, but are
solutions of a related classical
field theory.

The most obvious discrepancy between 
our result \rf{infrared}
and the energy-entropy argument of
Kosterlitz and Thouless is that we find a critical coupling of 
$e_{0}=2\sqrt{2}=2.83$ instead of $e_{0}=\pi/2=1.57$. This
need not concern us, for this number is not universal. A recent 
Monte-Carlo computation \cite{hasenbusch}
for the square-lattice Villain model indicates a critical coupling of $e_{0}=1.33$. The numerical result 
certainly disagrees with both analytical estimates.

\section{River deltas and solitons in the $O(3)$ sigma-model}

\setcounter{equation}{0}
\renewcommand{\theequation}{11.\arabic{equation}}

Next let us turn to the $O(3)$ case. The extrema of \rf{5.1} are related to
solutions of the spherical pendulum 
(see for example Whittaker \cite{whittaker} and Appendix B in this article). 

The main new ingredient in the spherical pendulum is the presence
of a new conserved quantity. This quantity is the angular momentum
about the vertical axis. It is not possible for the
pendulum to reach the zenith or nadir on the sphere
unless the angular momentum is zero. This is easy to understand 
physically. If the pendulum is placed at either point, the
angular momentum about the vertical axis is zero unless
the velocity diverges; but if the velocity does diverge, the total energy would also 
be divergent. Therefore there is a minimum distance
from the zenith and another from the nadir to the vertical axis. We have just
proved that the motion always lies between two circles on the sphere, each
circle
perpendicular to the vertical axis. Since we already know that
the vertical component of velocity
is periodic in time (in fact an elliptic function) it follows that the
pendulum actually touches the circles alternately and that the time
to go from one circle to the other is half the period. It can be proved
that the azimuthal angle always advances each half-period by 
$\Delta \kappa$, which
satisfies the Halphen inequality $\vert \Delta \kappa \vert \le \pi$ and the
Puiseux inequality $\vert \Delta \kappa \vert \ge \pi/2$ (both of these inequalities can be proved
using elementary complex analysis \cite{weinstein}). An illustration of the
pendulum motion is given in Figure 6.

\begin{figure}[t]
\centerline{\epsfxsize=3.5in \epsfysize=3.5in
\epsfbox{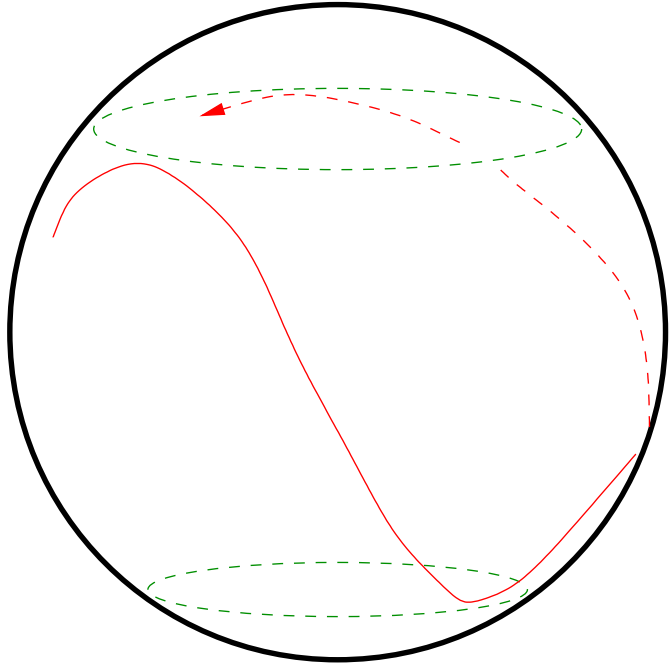}}
\label{6}
\caption[l]{\it The path of the spherical pendulum (shown in red). The motion is restricted
between two circles (shown in green). The advance in 
azimuthal angle as the pendulum moves from one circle
to the other is between $\pi/2$ and $\pi$. This also depicts an extremum of \rf{5.1}
for the $O(3)$ sigma model, where the role of time is played by the 
position coordinate $x$.}
\end{figure}

The configuration which extremizes the functional \rf{5.1} 
can be written immediately once the spherical
pendulum solution is known. It is most compactly written in terms of 
Weierstrass functions (our conventions are those of Whittaker and Watson \cite{WW})
${\cal P}(z)={\cal P}(z\,;\,g_{2}, g_{3})$ and their integrals. The 
period $2\omega_{1}$ is purely real, while the period $2\omega_{3}$ is purely
imaginary. Explicitly, the analysis of Appendix B gives
\beq
\xi(x)=\cos^{-1}\left[ -\frac{{\cal E}}{3}-{\cal P}
(y+\omega_{3}) \right] \;, \label{spherical1}
\eeq
\beq
e^{i\,\kappa(x)}
                &=&\exp\{
(\eta_{1}+\eta_{2})(\nu_{+}-\nu_{-})-[\zeta(\nu_{+})-\zeta(\nu_{-})]y \} \nonumber
\\      \nonumber  \\
                &\times&
{\sqrt {\frac{\sigma(y+\omega_{3}+\nu_{+})\;\sigma(y+\omega_{3}-\nu_{-})}
{\sigma(y+\omega_{3}-\nu_{+})\;\sigma(y+\omega_{3}+\nu_{-})}}} \;, \nonumber \\
\label{spherical2}
\eeq
where
\beq
y=\frac{2N\omega_{1}}{L}(x-x_{0}) \;. \nonumber
\eeq
There 
are two independent real parameters in this solution, because the 
constants in (\ref{spherical1}),(\ref{spherical2}) are related
by 
\beq
\frac{g_{2}}{4}=1+
\frac{{\cal E}^{2}}{3}\;,\;\; 
\frac{g_{3}}{4}=\frac{2\EE^{3}}{27}-\frac{2\EE}{3}+b\;, \nonumber 
\eeq
\beq
\nu_{+}=\omega_{1}+i\beta\;,\;\; 
\nu_{-}=i\gamma\;,\nonumber 
\eeq
\beq
{\cal P}(\nu_{\pm})=-\frac{{\cal E}}{3}\pm 1\;,\;\; \WP^{\r}(\nu_{\pm})=-2i{\sqrt b}\;, \label{parameters}
\eeq
$g_{2}$ and $g_{3}$ are related to $\omega_{1}$ and $\omega_{3}$ 
in the standard way and $\beta$ and $\gamma$ are real positive numbers, which
are less than $-2i\omega_{3}$. This 
is the most general solution, but it usually does not
satisfy the periodic boundary condition. The
expression (\ref{spherical1}) is periodic in $\xi$, that is  $\xi(L)=\xi(0)$, but in general
(\ref{spherical2}) is not periodic in $\kappa$.

We note that the invariants $g_{2}$ and $g_{3}$ are not completely arbitrary, because
of the inequalities
\beq
e_{1}>1-\frac{\EE}{3}\;,\;\;-1-\frac{\EE}{3}\le e_{3} < e_{2}\le 1-\frac{\EE}{3}\;. \label{ineq}
\eeq

From (\ref{spherical2}) it is found that the advance in $\kappa$ during a half-period is
\beq
\Delta \kappa=\kappa(L)-\kappa(0)=2i\omega_{1}[\zeta(\nu_{+})-\zeta(\nu_{-})]
-2i\eta_{1}(\nu_{+}-\nu_{-})\;. \nonumber
\eeq

Imposing periodic
boundary conditions forces 
$\Delta\kappa$ to be a rational multiple of $\pi$. The  
possibilities
include  $\Delta \kappa=\pm \pi$. Then the spherical pendulum solution reduces to that
of the planar pendulum with the identification 
$\sin \alpha =\sin \xi
\sin \kappa$
and $\cos \alpha=\cos\xi$. In this case, $U[\psi]$ coincides
with the $O(2)$ expression as does  
$\rho[\psi,\psi_{0}]$ (this follows immediately from \rf{3.3}).

The results of the previous paragraph 
show the conclusion of Feynman
\cite{feynman} that orbits containing configurations of 
nearly-constant derivatives are a small metric distance from $\psi_{0}$
is false. He considered the 
distance between orbits containing circulating configurations
such as \rf{soln2} and $\psi_{0}$ (his configurations were not explicit mathematically, but
he correctly argued that circulating configurations could be made to have arbitrarily
small potential energy). The distance $\rho^{circ}(k)$ we have already calculated shows
that 
these are far 
(of order $\sqrt L$) from $\psi_{0}$ for lowest potential energy. Feynman
claimed that a path in orbit space between these two 
orbits could be 
made short by what he called 
``slipping the loops about". He was
referring to the fact that the first 
homotopy group of the two-sphere is
trivial. However, this statement 
is false because the triangle inequality \rf{tri}
implies any such path has a length greater than $\rho^{circ}(k)$
\cite{alex}. Nonetheless, there is some merit
in Feynman's notion of making paths 
by ``slipping the loops about". We 
will show later in this section that most of the
important tunneling paths are of this type.

The configurations \rf{spherical1}, \rf{spherical2} for the $O(3)$ sigma model are related to
the soliton configurations
of a closely related model. Let us consider
adding an external source 
to the $O(3)$ sigma model action \rf{action}
as we did for the $O(2)$ model in \rf{sine-gordon}:
\beq
S=\frac{1}{2e_{0}}\int dt\,dx\, 
(\,\partial_{t} s^{T}\partial_{t} s
-{\partial}_{x} s^{T}{\partial}_{x} s+ h^{T}s\,
) \;. \nonumber
\eeq
This is no longer the periodic sine-Gordon action, as there
are two independent field components, instead of 
one. The direction of $h$ can be chosen 
so that the action becomes
\beq
S=\frac{1}{2e_{0}}\int dt\,dx\,  
\{ (\partial_{t} \xi)^{2}
-(\partial_{x} \xi)^{2}
+\sin^{2}\xi\, [(\partial_{t} \kappa)^{2}
-(\partial_{x} \kappa)^{2}]
+h\,\cos\xi 
\} \;, \nonumber
\eeq
and the equation of motion is 
\beq
(\partial_{t}^{2}-\partial_{x}^{2}) \xi 
+(\partial_{t}\sin^{2}\xi \,\partial_{t}
-\partial_{x}\sin^{2}\xi \,\partial_{x})\kappa=-h\sin\xi \label{sgeq2}\;.
\eeq
Now substituting $\xi(t,x)=\xi(u)$, $\kappa(t,x)=\kappa(u)$ where
$u=t\pm\frac{x}{v}$, as before, into 
\rf{sgeq2} yields the equations of motion
of the spherical pendulum:
\beq
\partial_{u}^{2}\xi(u)+\partial_{u}(\sin^{2}\xi \,\partial_{u}\kappa)
 =-\frac{v^{2}h}{(v^{2}-1)} \sin\xi \;, \nonumber
\eeq
The solutions
to this equation are a train of solitons, just as in the
$O(2)$ case, with spacing
\beq
l={\sqrt \frac{32 K^{2} (v^{2}-1) }{v^{2}h}  } \;. \nonumber 
\eeq
The difference is that this train is
not, in general, a periodic configuration. The solitons change from
one to the next, because of the angular shift $2\Delta \kappa$. Periodic
soliton solutions do exist when $\Delta \kappa=\pi$, but these
are a small subclass. While \rf{sgeq} is
a completely integrable partial differential equation, we do not expect that \rf{sgeq2}
is integrable. Furthermore, there is no notion of
topological stability for the solutions of the
$O(3)$ equation \rf{sgeq2}.

From the discussion above and Appendix A, it is 
clear that the $O(2)$ circulating solutions \rf{soln2} is
a local minimum of the potential energy, as is the $N=1$ oscillating solution
for small $k$, when periodic boundary conditions are imposed.

It is not sufficient to examine only the $O(2)$ solutions \rf{soln1} and \rf{soln2}
embedded in the $O(3)$ model. The reason is that \rf{spherical1} and \rf{spherical2}
contain a much larger class of highly nonlinear 
minimal-energy 
configurations, which should
be just as important in the thermodynamic limit. How do we impose the
boundary condition? We will answer this question after examining first
some general properties of \rf{spherical1} and \rf{spherical2}.

For the time being we assume no special boundary condition, but a very large
system, where $\cos \xi$ has period $l=L/M$, which is large compared to the short-distance
cut-off. Then we can see that integrating
over the long distance $L$ should give a nonvanishing result for 
$\int_{0}^{L} s_{3}dx/N$, but
zero for $\int_{0}^{L} s_{1}dx/N$ and $\int_{0}^{L} s_{2}dx/N$. This is an
ergodicity argument, and the result seems obvious, but is probably difficult to 
prove. Then the distance from $\psi_{0}$ 
is given by
\beq
{\rho(N;g_{2},g_{3})^{2}}=L-\vert \int_{0}^{L}\cos\xi\, dx \vert
=(1+\frac{\EE}{3})L-\frac{L}{2N\omega_{1}}[\zeta(2N\omega_{1}+\omega_{3})-\zeta(\omega_{3})] \nonumber
\eeq
\beq
=L\left( 1+\frac{\EE}{3}-\frac{\eta_{1}}{\omega_{1}}\right)    \label{o3dis}
\eeq
The 
potential
energy
can be written down as an integral:
\beq
U(N;g_{2},g_{3})=
\frac{8N^{2}\omega_{1}^{2}}{L^{2}}
\int_{0}^{L}\left\{ (1+\EE)-\left[ 1+\frac{\EE}{3}
+\WP \left( \frac{2N\omega_{1}x}{L}+\omega_{3}\right) 
  \right] \right\}\, dx\;.\nonumber
\eeq
This integral can be evaluated and we find
\beq
U(N;g_{2},g_{3})=\frac{8N^{2}\omega_{1}^{2}}{L^{2}} \left( \frac{2}{3}\EE
+\frac{\eta_{1}}{\omega_{1}} \right) \;.\label{o3pe}
\eeq

We know that for some choice of invariants 
$g_{2}$ and $g_{3}$, the right-hand 
side of \rf{o3pe} will diverge in the ultraviolet,
since it must reduce to the $O(2)$ result when the angular-momentum 
parameter $b$ vanishes. Examining the integral expression for
$\omega_{1}$:  
\beq \omega_{1}=\int_{e_{1}}^{\infty} [4 (z-e_{1})  
(z-e_{2})  (z-e_{3})]^{-\frac{1}{2}}\,dz \;, \nonumber 
\eeq it
is clear that it diverges only when $e_{1}-e_{2} \rightarrow 0$. On the other 
hand, $\eta_{1}/\omega_{1}$ does not diverge anywhere
except as $e_{1}\rightarrow \infty$, in which case \rf{o3pe} vanishes. This 
can be seen by writing this expression in term of
Jacobi elliptic integrals as 
\beq \frac{\eta_{1}}{\omega_{1}}=
-e_{1}+(e_{1}-e_{3})\frac{E(k)}{K(k)} \;, \label{elliprel1} 
\eeq
where 
\beq k^{2}=\frac{e_{2}-e_{3}}{e_{1}-e_{3}} \;. \label{elliprel2} \eeq 
In fact, we can also do this for $\omega_{1}$:  
\beq
\omega_{1}=\frac{K(k)}{\sqrt{e_{1}-e_{3}}} \;. \label{elliprel3} \eeq 
The relation \rf{elliprel3} can be obtained from the
identity 
\beq \WP(z)=e_{3}+\frac{e_{1}-e_{3}} {sn^{2}((e_{1}-e_{3})^{1/2}z,k) }\;, \label{elliprel4}
\eeq 
and identifying the periods, where 
$k$ is given by \rf{elliprel2}. Integration of \rf{elliprel4} and substitution of
$\zeta(\omega_{1})=\eta_{1}$ yields
\rf{elliprel1}. The potential energy diverges logarithmically 
when $e_{1}-e_{2}\rightarrow 0$, or $k\rightarrow 1$. It is otherwise
finite, vanishing in the thermodynamic limit.

To understand better where $U(N;g_{2},g_{3})$ diverges, we write
\beq
e_{1}=W+\delta\;,\;\; e_{2}=W-\delta\;,\;\; e_{3}= -2W \;.\nonumber
\eeq
If $\delta<1/4$, the inequalities \rf{ineq} become
\beq
\frac{2+\delta}{3}
>W>\frac{2-\delta}{3}   \;. \label{newineq}
\eeq
These new inequalities are completely consistent with the relations \rf{parameters}. The 
divergence occurs as $\delta \rightarrow 0$. We will compute the angular-momentum parameter
$b$ and the azimuthal-angle advance $\Delta \kappa$ in this limit.

\begin{figure}[t]
\centerline{\epsfxsize=3.5in \epsfysize=3.5in
\epsfbox{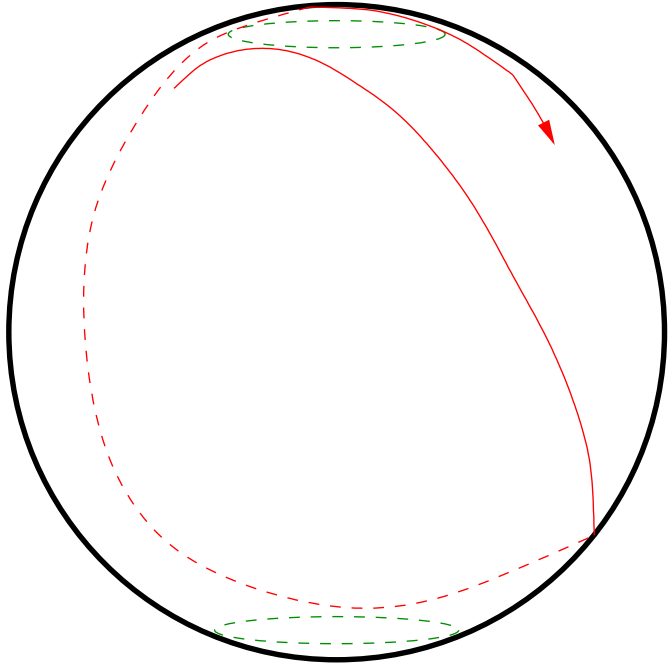}}
\label{7}
\caption[l]{\it The configurations (11.1), (11.2)
with large potential energy are spherical pendulum solutions
for which the circles begin to close around the 
vertical axis. As the circles close completely, the 
advance in azimuthal
angle approaches $\pi$ and the
dynamics of the spherical pendulum 
becomes the
dynamics of the planar pendulum.}
\end{figure}

As $\delta$ vanishes, $e_{1}$, $e_{2}$ and $e_{3}$ approach $2/3$, $2/3$ and 
$-4/3$, respectively, by \rf{newineq}. We also find that
$\EE \rightarrow 1$. Then ${\cal P}(\nu_{\pm})=-\frac{{\cal E}}{3}\pm 1$ implies
that 
$\WP(\nu_{+})=2/3$ and $\WP(\nu_{-})=-4/3$. Hence $\nu_{+}=\omega_{1}$ and $\nu_{-}=\omega_{3}$. But then 
$\WP^{\r}(\nu_{\pm})^{2} =0$, so that $b=0$. It must therefore be true that a
configuration in the orbit 
becomes
an $O(2)$ extremal-curve configuration with $k \approx 1$. This can be checked
by seeing whether
\begin{itemize}
\item $\vert \Delta \kappa \vert$ is $\pi$, and
\item $\rho(N;g_{2},g_{3})$ approaches zero as $e_{1}-e_{2} \rightarrow 0$.
\end{itemize}
Explicitly
\beq
\Delta \kappa \rightarrow 2i
\omega_{1}[\zeta(\omega_{1})-\zeta(\omega_{3})]-2i\eta_{1}(\omega_{1}-\omega_{3})
=2i(\eta_{1}\omega_{2}-\eta_{2}\omega_{1})=\pi \;.\nonumber
\eeq
From \rf{o3dis}, \rf{elliprel1} and \rf{newineq},
\beq
{\rho(N;g_{2},g_{3})^{2}}\rightarrow
L\left( 1+\frac{1}{3}-\frac{2}{3} \right) =0 \;.   \nonumber
\eeq

\begin{figure}[t]
\centerline{\epsfxsize=4.5in \epsfysize=3.5in
\epsfbox{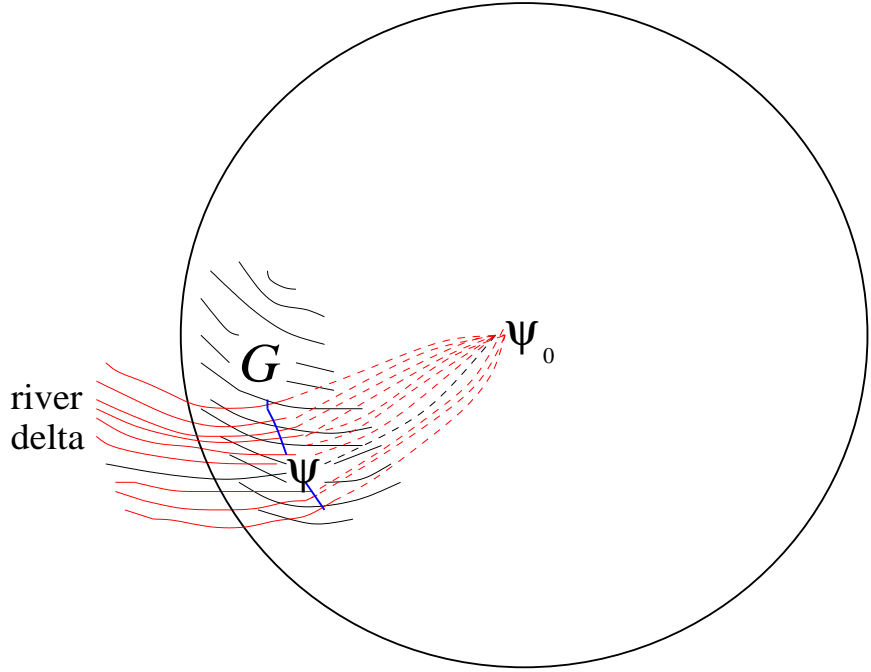}}
\label{8}
\caption[l]{\it A two-dimensional river delta where tunneling can occur
most readily though
ultraviolet-divergent energy
barriers. The $O(2)$ river valley passing through the point $\psi$
in the orbit-space sphere 
around $\psi_{0}$ is drawn in black, while the remainder
of the delta is drawn in red. The curve of intersection of the 
delta and the sphere is drawn in blue.}
\end{figure}

This shows that as the potential energy diverges, the distance between
the configurations \rf{spherical1}, \rf{spherical2} and $\psi_{0}$ vanishes. This property
was also shown to be true for the $O(2)$ model. What is also
significant is that in this limit the configuration becomes
periodic! There are potential-energy barriers
just as in the $O(2)$ model. In fact, the orbits close to
$\psi_{0}$
become what is essentially the $O(2)$ configuration for $k\approx 1$. The situation is
shown schematically in Figure 7.

Imagine
moving along the barrier from $\psi_{0}$. As $\rho$
increases and the energy decreases the parameter $e_{1}$ becomes more
important, as it is no longer forced to be $2/3$. There is
now an {\it extremal fan} 
consisting 
of many adjacent 
extremal curves (with $e_{1}-e_{2}$ small), converging 
at the point of orbit space $\psi_{0}$. Consider a neighborhood {\it
G} in
the sphere around a $b=0$ solution $\psi$ of the 
form \rf{spherical1}, \rf{spherical2} (this is identical 
to an
$O(2)$ extremal-curve solution). There is a 
one-dimensional curve containing $\psi$ where the fan intersects the
sphere. The endpoints of this curve are fixed by \rf{newineq}. For
any $N$, the fan contains a circulating $O(2)$ river valley \rf{soln1}. We
call the fan a river delta. The potential
energy on this river delta is less than on the rest of {\it G}. The amplitude
for barrier penetration is greatest along the delta. The situation is
depicted in Figure 8.

The paths along the river delta are essentially Feynman's
``slipping the loops about" paths. The $O(2)$ river valley orbits inside
the $O(3)$ river delta contain very symmetrical configurations, which
are preserved, up to a sign, under reflection $x\rightarrow L-x$. This
is not so for the general river-delta configurations. The first
homotopy group of $O(3)$ is trivial and it is not hard to visualize
these paths as the decay of a non-topologically-stable soliton.

Since the extremal-fan configurations with significant
potential energy are periodic, we make the approximation that {\it all} 
river-delta configurations can be replaced with modified configurations
which satisfy periodic boundary conditions. We add
an extra linear term to $\kappa(x)$, resulting in 
these
boundary conditions. These are not strict
minima of the potential energy for a given $\rho$. However their
potential energy 
is the same as that of the corresponding $O(2)$ extremal-curve configurations 
within order $1/L$. The modification we make is to replace
\rf{spherical2} by
\beq
e^{i\,\kappa(x)}
&=&\exp\{
                (\eta_{1}+\eta_{2})(\nu_{+}-\nu_{-})-[\zeta(\nu_{+})-\zeta(\nu_{-})]y  \nonumber \\
&+&\frac{\pi}{2\omega_{1}}y
         +2[\zeta(\nu_{+})-\zeta(\nu_{-})]y
         -2\eta_{1}(\nu_{+}-\nu_{-})y \} \nonumber  \\
&\times&
{\sqrt {\frac{\sigma(y+\omega_{3}+\nu_{+})\;\sigma(y+\omega_{3}-\nu_{-})}
{\sigma(y+\omega_{3}-\nu_{+})\;\sigma(y+\omega_{3}+\nu_{-})}}} \;. \;\;\;\;\;\;\;\;\;\;\;\;\;
\;\;\;\;\;\;\;\;\;\;\;\;\;\;\;\;\;\;\;\;\;\;\;\;\;\;\;\;\;\;\;\;\;\;\;\;\;\;\; 
\label*{\rf{spherical2}^{\r}} \nonumber
\eeq
It is now permissible to consider such a configuration
with any choice of $N$. The river deltas in
the model with periodic boundary conditions are paths
through configurations of the form \rf{spherical1}, \rf{spherical2}$^{\r}$.

\section{Tunneling and the topological charge}

\setcounter{equation}{0}
\renewcommand{\theequation}{12.\arabic{equation}}

Adding the term
\beq
S_{\theta}&=&\th \,q = \frac{\theta}{4\pi} \int dt \,dx \;\epsilon^{a\,b\,c}\, s_{a}\,\partial_{t}s_{b}\,
            \partial_{x}s_{c}  \nonumber \\
       &=&\frac{1}{4\pi}\int dt \,dx \;[\,
\partial_{t} (\cos\xi\, \partial_{x} \kappa) -\partial_{x} (\cos\xi \,\partial_{t} \kappa) \,] \;.\nonumber
\eeq
to the action has a significant effect on tunneling through the energy
barriers. If the spacetime is the two-sphere and
the action of the spacetime configuration $s(t,x)$
is finite, then the number $q$ is an integer, called the 
topological charge, and
is the degree of the mapping from the two-sphere to itself. If
the action is not finite, $q$ is not quantized and
the term ``topological charge" is 
a misnomer. We 
will show in this section that the space-time field
configuration generated by moving from
$\psi_{0}$ through the $N=1$
river delta has a half-integer value of $q$.

As in the $O(2)$ model, an ultraviolet cut-off must be introduced. The
tunneling for $N=1$ proceeds as follows: \begin{enumerate}
\item First a wave packet of the form
\rf{wavepacket} expands freely
from a constant configuration in 
$\psi_{0}$ to an $N=1$ extremal-fan configuration whose derivatives are limited 
by the
cut-off. 
\item Next this river-delta configuration
begins to stretch out continuously in space (its derivatives with
respect to $x$ becoming
smaller) as
$\rho$
increases. Eventually the configuration becomes
a constant (in $\psi_{0}$)
once again. 
\end{enumerate}
Since 
boundary conditions in $x$ are periodic, Stokes' theorem implies
\beq
q=
\lim_{e_{1}-e_{2}\rightarrow 0} \;\frac{1}{4\pi} 
\int_{0}^{L} \cos\xi\, \partial_{x} \kappa  \;.    
    \label{degree}
\eeq

Instead of doing a detailed calculation of the degree using our explicit
configurations \rf{spherical1}, \rf{spherical2}$^{\r}$, we can see what
\rf{degree} must
be from an intuitive
argument. The azimuthal-angle advance in the initial river-delta configuration must approach
$\Delta \kappa =\pi$ as $e_{1}-e_{2}$ approaches zero, as we have already shown. This advance
takes place over extremely short intervals in $x$. Since $N=1$, the advance happens
once on the pendulum down-swing $x=x_{0}+L/2$, and again on
the up-swing at 
$x=x_{0}$. Since $\kappa$ is a compact variable, we must subtract $2\pi$ from the
advance at the up-swing. As $e_{1}-e_{2} \rightarrow 0$
\beq
\kappa \rightarrow \pi [H(x-x_{0})-H(x-x_{0}-L/2)] \;, \nonumber
\eeq
where $H$ is the step function, namely $H(x)=0$ for $x<0$, $H(x)=1$ for $x\ge 0$. Thus \rf{degree} is
\beq
q&=&\frac{1}{4\pi}\int_{0}^{L} \cos\xi(x) \cdot \pi\;
    [\delta(x-x_{0})-\delta(x-x_{0}-L/2)]\;dx   \nonumber \\
 &=&\frac{1}{4}[\cos\xi(x_{0})-\cos\xi(x_{0}+L/2)] \;.\nonumber
\eeq
Since $b\rightarrow 0$, the pendulum almost swings through the zenith and the nadir, so that
$\xi(x_{0})\rightarrow 0$ and $\xi(x_{0}+L/2)\rightarrow \pi$. The final result is
\beq
q=\frac{1}{4}[1-(-1)] =\frac{1}{2} \;. \nonumber
\eeq

As an orbit moves
along the path through the barrier 
joining $\psi_{0}$ to itself, the wave functional must pick up
a phase $\exp \pm i\th/2$. At $\theta=\pi$, passage through the barriers contribute 
factors of oscillating sign to the wave functional. This means that the
wave functional will vanish somewhere along the barrier. Barrier configurations
with $N=1$ are suppressed. However, multi-soliton barrier configurations may survive.

\section{Discussion}

Our Hamiltonian methods, though semiclassical in nature, are very
different from the saddle-point techniques most field theorists
are accustomed to. They are conceptually more complicated than saddle-point
methods, but are at least
partially successful for the $O(3)$ sigma model. We
do not know yet whether they can be developed to obtain quantitative results
or if they might suggest a more powerful version of the 
saddle-point methods.

We have not 
evaluated the barrier-penetration amplitude in 
the $O(3)$ sigma model. This is a formidable
problem for several reasons. The regions of barrier penetration in orbit
space
are two-dimensional, unlike the case of the $XY$ model, for which
they are one-dimensional. More significantly, the task of
understanding
the role of nonlinear spin-wave
fluctuations near the barriers will probably not be easy. In the $XY$
model, further spin-wave corrections are not important because
the spin waves are effectively
Gaussian. 

Though there is still much we do not
understand about the $O(3)$ model, our analysis has been 
revealing. We have found the
explicit form of the barrier configurations. These truly have half-integer
topological charge, just as predicted by Affleck and Haldane 
\cite{AffHal}. However, the one-parameter family of barrier configurations are more 
general and more complicated than both the modified
meron in Affleck's paper and the original meron of Gross \cite{gross}.

When $\theta=0$, the Schr\"{o}dinger wave functional on
configuration space can
be significantly different from zero only where the potential
energy is small or near a river-delta configuration. This is strong
evidence that barrier penetration produces the mass gap. The
fact that the topological charge is half-integer means that at $\theta=\pi$, there
can be 
a massless phase. A phase transition may separate 
this from a massive strong-coupling phase, driven by pairs of 
barrier-penetration events (of vanishing or integer topological charge).

Feynman's \cite{feynman} point of view as to how the gap arises was rather
different from ours. He argued that the mass gap should arise because the vacuum
wave functional should vanish everywhere on the orbit space, except on a region
of finite diameter. This may be true, though it is a subtle matter in
an asymptotically-free theory where all length scales are important. Quantum
barrier penetration will occur for configurations other than the
classical vacuum $\psi_{0}$; one effect of such barrier penetration may be
to suppress configurations which are a distance $\rho \sim \sqrt L$
from $\psi_{0}$.

We have left many questions unanswered and
see three further directions for this research.

While we find the
phase transition in the Hamiltonian $XY$ model, our technique is
not yet powerful enough to fully understand even this model. Our 
methods are as good as 
vortex-mean-field theory \cite{kt}, but we are not yet able to
do a renormalization group analysis, as Kosterlitz first did 
for the Coulomb gas (or its equivalent, the sine-Gordon model). It
may be that a Hamiltonian renormalization group
of the barrier configurations is possible.

We would like to understand better the condensation
of barrier configurations in the $O(3)$ model. This
will require a proper incorporation of
spin-wave effects. In any case, our results
indicate that the instanton gas
of integer topological charges \cite{instantons} is
not sufficient to understand mass generation
in the model. The one-parameter
configurations
we have found appear to be more general 
than merons \cite{gross}. We expect that
if a calculation could be done by a saddle-point
method in Euclidean space, it
would also reveal this to be the case.

Finally, the extremal problem for the sigma model \rf{5.1} has
an analogue in gauge theories. Some progress has been made
towards its solution \cite{orland1}, building on some
of the work here and in reference \cite{orland}.

\section*{Acknowledgements} P.O. thanks the Niels Bohr Insitute staff for
their hospitality. We would also like to thank W. Bietenholz for a careful
reading of the manuscript.

\section*{Appendix A: Which extremal curves are minimal curves?}

\setcounter{equation}{0}
\renewcommand{\theequation}{A\arabic{equation}}

In this appendix we determine which of our extremal curves for the
$O(2)$ model are
what we call river valleys. These
must be minimal curves. We need to know whether a point (actually
a field configuration or an orbit) on such a curve, constrained to
be on the sphere (of constant $r$ from $s_{0}$ or of constant $\rho$
from $\psi_{0}$) is in stable classical equilibrium.

Once the situation is understood for the $O(2)$ model, it
is possible to decide the issue of river deltas in the $O(3)$
model, discussed in section 11.

Consider a solution of \rf{5.1} for the $O(2)$ 
case. Suppose that the angle $\alpha(x)$ is an extrema curve configurations
such as \rf{soln1} or \rf{soln2}. If we vary this angle by $\delta \alpha(x)$, the 
third
potential energy changes to second order in
$\delta \alpha$ by
\beq
\delta U
&=&-\int_{0}^{L} \left( 
\delta \alpha \frac{d^{2}}{dx^{2}} \alpha
+\frac{1}{2} \delta \alpha 
\frac{d^{2}}{dx^{2}} \delta \alpha 
\right)\;dx     \nonumber \\
&=&-\frac{1}{2} \int_{0}^{L} \;\delta \alpha \left(\frac{d^{2}}{dx^{2}} \delta \alpha 
-2\lambda \sin \alpha
\right) 
\;dx
\;, \label{varen}
\eeq
where $\lambda$ is the Lagrange multiplier in
\rf{5.1} and we have used the pendulum equation of motion. We 
want to know whether $\delta U$ is positive for any acceptable 
$\delta \alpha$, other than the zero mode 
of 
translation 
invariance. We say ``acceptable" because $\delta \alpha$ is not
arbitrary. There is a
constraint in \rf{5.1} on $\delta\alpha$, namely that
the variation must not change the distance in orbit space
to $\psi_{0}$.

The constraint means that $\delta \alpha$ must satisfy
\beq
v^{2}
&=&\left[ \int_{0}^{L} \;\cos(\alpha+\delta \alpha) \;dx \right]^{2}
   + \left[ \int_{0}^{L} \;\sin(\alpha+\delta \alpha) \;dx \right]^{2} \nonumber \\
&=&\left( \int_{0}^{L} \;\cos\alpha \;dx \right)^{2}
+ \left( \int_{0}^{L} \;\sin\alpha \;dx \right)^{2}
\;.          \nonumber
\eeq
Expanding this  up to second order in $\delta \alpha$ gives
\beq
0&=&-v_{1}\int_{0}^{L} \left[ \delta \alpha \,\sin \alpha+\frac{1}{2}(\delta \alpha)^{2} \,\cos \alpha \right]\;dx
+\left( \int_{0}^{L} \delta \alpha\, \sin \alpha\;dx \right)^{2} \nonumber \\
&+&v_{2}\int_{0}^{L} \left[ \delta \alpha \,\cos \alpha-\frac{1}{2}(\delta \alpha)^{2} \,\sin \alpha \right]\;dx
+\left( \int_{0}^{L} \delta \alpha\, \cos \alpha\;dx \right)^{2} \;, \label{varconstr}
\eeq
where 
\beq
v_{1}=\int_{0}^{L} \cos \alpha \;dx \;,\;\; 
v_{2}=\int_{0}^{L} \sin \alpha \;dx \; . \nonumber
\eeq
The second-order form of the constraint \rf{varconstr} can be 
simplified somewhat.

Let us rotate $\alpha$ so that $v_{1}=v$ and $v_{2}=0$, as is the case for \rf{soln1}, \rf{soln2}. The same conditions can be
imposed
on $\alpha+\delta \alpha$. The expansion of these conditions to second order yields
\beq
0&=&\int_{0}^{L} \left[ \delta \alpha \,\sin \alpha+\frac{1}{2}(\delta \alpha)^{2} \,\cos \alpha \right]\;dx \label{newvarc1} \\
0&=&\int_{0}^{L} \left[ \delta \alpha \,\cos \alpha-\frac{1}{2}(\delta \alpha)^{2} \,\sin \alpha \right]\;dx\;. \label{newvarc2}
\eeq
Conditions \rf{newvarc1} and \rf{newvarc2} are equivalent to \rf{varconstr}. The variation of the potential
energy \rf{varen} upon substitution of \rf{newvarc1} is
\beq
\delta U
=-\frac{1}{2} \int_{0}^{L} \;\delta \alpha \left(\frac{d^{2}}{dx^{2}} 
+\lambda \cos \alpha 
\right) \delta \alpha 
\;dx
\;, \label{varen1}
\eeq
Substituting \rf{soln1} and \rf{soln2}, equation \rf{varen1} becomes
\beq
\delta U
=\frac{1}{2} \int_{0}^{L} \;\delta \alpha \left[-\frac{d^{2}}{dx^{2}} 
+\frac{16N^{2}K^{2}}{L^{2}} 
\left(-1+2k^{2}\,sn^{2}  
\frac{4NK(x-x_{0})}{L} 
\right)
\right] \delta \alpha 
\;dx  \label{lame1}
\eeq
and
\beq
\delta U
=\frac{1}{2} \int_{0}^{L} \;\delta \alpha \left[-\frac{d^{2}}{dx^{2}} 
+\frac{4N^{2}K^{2}k^{2}}{L^{2}} \left(-1+2\,sn^{2}  \frac{2NK(x-x_{0})}{L} \right)
\right] \delta \alpha 
\;dx 
\;, \label{lame2}
\eeq
respectively. 

The eigenvalue equation for the
operator in square brackets in each of \rf{lame1} and \rf{lame2} is a Hill equation, specifically
a Lam\'{e} equation 
\cite{WW} of order one. It is still necessary to impose the constraints \rf{newvarc1}
and \rf{newvarc2}.

Our problem has become the following question: under what circumstances
are the quadratic forms
in \rf{lame1} and \rf{lame2} positive on variations satisfying
\rf{newvarc1} and \rf{newvarc2}? 

After an appropriate rescaling, the Lam\'{e} operators are
\beq
{\cal H}_{osc}
={\cal H}-1   \;, \label{hill1} 
\eeq
on periodic functions of period $4NK$, for 
the oscillating extremal curves \rf{soln1} and
\beq
{\cal H}_{circ}={\cal H}-k^{2} \;, \label{hill2}
\eeq
on periodic
functions of period $2NK$, for 
the circulating extremal curves \rf{soln2}, where in each case
\beq
{\cal H}=-\frac{d^{2}}{du^{2}}+2k^{2}\,sn^{2}\,u  \;. \nonumber
\eeq
To answer this 
question it is first necessary to see whether the spectrum of
\rf{hill1} or \rf{hill2} is positive-definite
with a single zero mode (corresponding to translation invariance
of the extremal curve configuration \rf{soln1} or \rf{soln2}, respectively). If
this is the case for a particular Lam\'{e} operator, any variation raises
the potential energy. If not, meaning that there is a negative eigenvalue
in the spectrum, it must then be
checked on a case-by-case basis
whether variations, other than the zero mode of translation invariance, satisfying 
\rf{newvarc1} and \rf{newvarc2} can give
a nonpositive $\delta U$.

The zero mode corresponding to translations for \rf{hill1} is $cn\,u$ and for \rf{hill2}
is $dn\,u$, which are both doubly-periodic
functions. These modes are obtained by simply differentiating \rf{soln1}
and \rf{soln2} with respect to $x_{0}$. Recall that the ground-state eigenfunction of a
quantum-mechanical Hamiltonian must
be a unique (i.e. nondegenerate) real 
function which vanishes nowhere in the physical region of $u$. Every other real eigenfunction
must have at least one node. In fact if a given eigenfunction possesses no zeros in
the physical region it must be the ground-state eigenfunction. This eigenfunction
is $dn\,u$ (which satisfies the boundary conditions for \rf{hill1} and \rf{hill2}).

Let us first consider the case of the oscillating extremal curves, \rf{hill1}. We 
can see readily that there is a negative eigenvalue in the spectrum. We have
shown that the ground-state eigenfunction for either
the oscillating and the circulating
case, is $dn\,u$ which has period $2K$ and no nodes. Its eigenvalue
of ${\cal H}_{osc}$ is $-k^{\r\,2}$. We will not do the analysis here to
determine the effect of the constraints \rf{newvarc1}, \rf{newvarc2}
on $\delta U$ (though it can probably be done, as the spectrum
of the order-one Lam\'{e} with our boundary conditions can be completely determined). We will
only note that the $N=1$ oscillating solution \rf{soln1} for sufficiently small $k$
{\it must} be a local minimum on the sphere of radius $\rho_{0}$. For
there are only three kinds of extremal configurations on the
sphere (see figure 1): \begin{itemize}
\item the $k<k^{*}$ oscillating solutions \rf{soln1}.
\item the $k>k^{*}$ oscillating solutions.
\item the circulating solutions \rf{soln2}, if $\rho^{circ}(0)\le \rho_{0}$.
\end{itemize}
If $\rho_{0}$ is sufficiently small, a configuration of the third type 
is a barrier configuration. In that case, among the three possibilities, the $N=1$
oscillating solution with $k<k^{*}$ has the smallest potential energy. Since the potential
energy is bounded below, it must have a minimum value on the sphere.

For the circulating extremal curves \rf{hill2}, the ground state $dn\,u$ is also the
zero mode, so the remainder of the spectrum is positive. Hence all the circulating
extremal curves are river valleys.

\section*{Appendix B: The spherical pendulum}
\setcounter{equation}{0}
\renewcommand{\theequation}{B\arabic{equation}}

The Lagrangian of the spherical pendulum is
\beq
L=\frac{1}{2} {\dot \xi}^{2}+\frac{1}{2} \sin^{2} \xi \,{\dot\kappa}^{2}
-\mu(1-\cos\xi)\;.  \nonumber
\eeq
The equation of motion for $\kappa$ is the statement that the vertical
component of the angular-momentum vector is conserved, that is
\beq
\frac{dl_{z}}{dt}=0\;,\;\;l_{z}=\sin^{2}\xi \;{\dot \kappa}\;. \label{C1}
\eeq
The conserved energy is 
\beq
E_{pend}=\frac{1}{2} {\dot \xi}^{2}+\frac{l_{z}^{2}}{2\sin^{2} \xi} 
+\mu(1-\cos\xi)\;. \;\;\;\;\;\;\;\;\;\;\;
\label{C2} 
\eeq
Define the new time coordinate $\tau={\sqrt\frac{\mu}{2}}t$ and the
new conserved quantities 
$b=\frac{l_{z}^{2}}{2\mu}$ and 
$\EE
=E_{pend}/\mu-1$. The energy-conservation relation, \rf{C2} becomes
\beq
\left( \frac{d\xi}{d\tau}\right)^{2}=
4(\EE+\cos\xi)-\frac{4b}{\sin^{2}\xi} \;.
\nonumber
\eeq
Let ${{\cal Z}}=-\cos\xi$. The energy-conservation equation is then
\beq
\left( \frac{d{{\cal Z}}}{d\tau}\right) ^{2}=
4{\cal Z}^{3}-4\EE{\cal Z}^{2}-4{\cal Z}+4(\EE-b)\equiv M(\ZZ) \;.
\label{C3}
\eeq
The physical region is, by definition, ${\cal Z}\in [-1,1]$. Since $(d{\cal Z}/d\tau)^{2}$
must be positive in this region, the cubic polynomial $M(\ZZ)$ on the
left-hand
side of 
\rf{C3}
must
have some positive values in this region. If ${\cal Z}=\pm 1$, then 
$M(\ZZ)$ has the value $-4b$. Hence the polynomial has at least two
real roots between $-1$ and $1$. The motion of the pendulum is such
that the value of $\ZZ$ is between these two 
roots. Furthermore, if \rf{C3} is continued
outside of the physical region, in the limit as ${\cal Z}\rightarrow
\infty$, the behavior of $M(\ZZ)$ is
$M(\ZZ)\rightarrow \infty$. Therefore
there is one (unphysical) root of $M(\ZZ)$ in the open interval ${\cal
Z}\in (1,\infty)$.

After an appropriate shift of $\cal Z$ by $-\EE/3$, eliminating 
the quadratic term in the
polynomial $M(\ZZ)$, it is straightforward to solve \rf{C3} in terms of the
Weierstrass 
elliptic function $\WP(z;g_{2},g_{3})=\WP(z)$. The solution is
\beq
\ZZ=\WP(\tau-\tau_{0})+\frac{\EE}{3}\, \nonumber 
\eeq
where
\beq
g_{2}=4\left( 1+\frac{\EE^{2}}{3}\right) \;,\;\;
g_{3}=4\left( \frac{2\EE^{3}}{27}-\frac{2\EE}{3}+b \right) \;. \nonumber
\eeq
We can parametrize the Weierstrass function by
$e_{1}$, $e_{2}$ and $e_{3}$ instead of $g_{2}$ and $g_{3}$. These numbers
are just the roots of the polynomial
on the right-hand side of \rf{C3} shifted by $-\EE/3$. They are real, add up to
zero and satisfy
\beq
e_{1}>1-\frac{\EE}{3}\;,\;\;-1-\frac{\EE}{3}\le e_{3} < e_{2}\le 1-\frac{\EE}{3}\;. \nonumber
\eeq
Since $e_{1}$ is positive, $e_{3}$ is negative, while $e_{2}$ can be of either sign. The
motion of the pendulum always lies between two circles on the sphere. The pendulum
reaches the bottom
circle at time $\tau$ such that
$\WP(\tau-\tau_{0})=e_{3}$ and the top circle at time $\tau$ such 
that $\WP(\tau-\tau_{0})=e_{2}$. If the 
initial condition at $\tau=0$ is chosen so that $\ZZ$ lies on the bottom circle, then
$-\tau_{0}$ is fixed to be $\omega_{3}$. 

Notice that
\beq
\omega_{1}=\int_{e_{1}}^{\infty} (4t^{3}-g_{2}t-g_{3})^{-1/2} dt  \nonumber
\eeq
is purely real, while
\beq
\omega_{3}=-i\int_{-\infty}^{e_{3}} (g_{3}+g_{2}t-4t^{3})^{-1/2} dt  \nonumber
\eeq
is purely imaginary.

Equation \rf{C1} can now be written
\beq
\frac{d\kappa}{d\tau}=\frac{2{\sqrt b}}
{1-[\WP(\tau+\omega_{3})+\frac{\EE}{3}]^{2}}\;. \label{C5}
\eeq
To integrate \rf{C5} requires two steps. The first is to reduce the denominator, which
is quadratic in $\WP(\tau+\omega_{3})$ to the sum of terms with denominators linear
in this function, that is
\beq
\frac{1}{1-(\WP+\frac{\EE}{3})^{2}}\;=\;
\frac{1}{2}\,\frac{1}{\WP+\frac{\EE}{3}+1}
-\frac{1}{2}\,\frac{1}{\WP+\frac{\EE}{3}-1} \;. \nonumber
\eeq
To integrate such an expression, we prove the identity
\beq
\frac{\WP^{\r}(\nu)}{\WP(z)-\WP(\nu)}=2\zeta(\nu)+\zeta(z-\nu)-\zeta(z+\nu)\;. \label{ident}
\eeq
We will show that each of the two sides of \rf{ident} has the same zeros and poles
(and multiplicities thereof) with the same residues of the poles. Equation \rf{ident} then
follows. Since $\zeta(z+2\omega_{i})=\zeta(z)+2\eta_{i}$, the right-hand side is an
elliptic function. The zeros of the right-hand side of \rf{ident} are points congruent to zero, since
the Weierstrass zeta function is odd. Furthermore, the order of
each of these zeros is two, for as $z\rightarrow 0$,
\beq
2\zeta(\nu)+\zeta(z-\nu)-\zeta(z+\nu) \rightarrow -z[\WP(-\nu)-\WP(\nu)]+O(z^{2})=O(z^{2})\;. \nonumber
\eeq
The poles of the right-hand side are points congruent to $z=\pm \nu$, since $\zeta(z)$ has poles
congruent to zero. These are simple 
poles. We therefore have in each period cell one zero
of order two and two poles of order one. The poles 
have residue $\pm 1$ at $\pm \nu$. Now the
left-hand side of \rf{ident} has double zeros at the double poles of 
$\WP(z)$; these are
the points congruent to zero. Furthermore, it 
has simple poles at the points congruent
to $\pm \nu$, with residue
\beq
\lim_{z\rightarrow \pm \nu} \frac{\WP^{\r}(\nu)}{\WP(z)-\WP(\nu)}\,(z\pm \nu)
=\pm 1\;. \nonumber
\eeq
Therefore \rf{ident} is correct. 

The right-hand side of \rf{ident} is easy to integrate, giving the result
\beq
\int \frac{dz}{\WP(z)-\WP(\nu)}
\;=\;\frac{1}{\WP^{\r}(\nu)} [2z \zeta (\nu)
+\log \frac{\sigma (z-\nu) }{ \sigma (z+\nu)}  ]\;. \nonumber
\eeq

Define two complex numbers $\nu_{+}$ and $\nu_{-}$ by 
\beq
\WP(\nu_{\pm})=-\frac{\EE}{3}\pm 1\;. \label{differ}
\eeq
These numbers are determined up to $2m\omega_{1}+2n\omega_{3}$ as well as an overall
sign. Now
\beq
\WP^{\r}(\WP^{-1}(z))=(4z^{3}-g_{2}z-g_{3})^{1/2} \;,\nonumber
\eeq
and substituting \rf{differ} gives
\beq
\WP^{\r}(\nu_{\pm})^{2}=-4b\;. \nonumber
\eeq
Since $\WP^{\r}$ is an odd function, we have the option of choosing
\beq
\WP^{\r}(\nu_{\pm})=-2i\,{\sqrt b}\;. \nonumber
\eeq
This choice is satisfied by the expressions for $\nu_{\pm}$
evaluated below.

Inverting \rf{differ} gives 
\beq
\nu_{\pm}=\omega_{3}+\int_{-\frac{\EE}{3}\pm 1}^{e_{3}}\frac{dz}{\sqrt{4\,
(z-e_{1})\,(z-e_{2})\,(z-e_{3})}} \;. \label{nupm}
\eeq
We can immediately see from \rf{nupm} and $-\frac{\EE}{3}-1<e_{3}<-\frac{\EE}{3}+1$
that $\nu_{-}$ is purely imaginary. On the other hand, $\nu_{+}$ has 
a nonzero real part. Explicitly,
\beq
\nu_{+}
=\omega_{3}-\int_{e_{3}}^{-\frac{\EE}{3}\pm 1}
\frac{dz}{\sqrt{4\,
(z-e_{1})\,(z-e_{2})\,(z-e_{3})}}  \nonumber 
\eeq
\beq
=\omega_{3}+\int_{e_{2}}^{-\frac{\EE}{3}+1}
\frac{dz}{\sqrt{-4\,(e_{1}-z)\,(z-e_{2})\,(z-e_{3})}} \nonumber
\eeq
\beq
-\int_{e_{3}}^{e_{2}}\frac{dz}{\sqrt{4\,(e_{1}-z)
\,(e_{2}-z)\,(z-e_{3})}}\;. \label{nup}
\eeq 
The first term of \rf{nup} is purely imaginary. So is
the second; this follows from $e_{1}>-\frac{\EE}{3}+1$. The last term 
is $\omega_{2}-\omega_{3}=\omega_{1}+2\omega_{2}$. We therefore have (adding
$-2\omega_{2}$ to $\nu_{+}$ which changes nothing)
\beq
\nu_{+}=\omega_{1}+i\beta \;,\;\; 
\nu_{-}=i\gamma\;, \nonumber
\eeq
where $\beta$ and $\gamma$ are real constants between zero and $-2i\omega_{3}$.

The solution of \rf{C5} can now be written down. It is
\beq
e^{i\,\kappa(x)}
                &=&\exp\{
(\eta_{1}+\eta_{2})(\nu_{+}-\nu_{-})-[\zeta(\nu_{+})-\zeta(\nu_{-})]\tau \} \nonumber
\\      \nonumber  \\
                &\times&
{\sqrt {\frac{\sigma(\tau+\omega_{3}+\nu_{+})\;\sigma(\tau+\omega_{3}-\nu_{-})}
{\sigma(\tau+\omega_{3}-\nu_{+})\;\sigma(\tau+\omega_{3}+\nu_{-})}}} \;. \label{azimuth}
\eeq

\vfill


\begin{thebibliography}{xx}
\bibitem{AF} A.~M. Polyakov, {Phys. Lett.} {\bf 59B} (1975) 79. 
\bibitem{INST} A.~A. Belavin and A.~M. Polyakov, {J.E.T.P. Lett.} {\bf 22} (1975) 245/
\bibitem{SC} F.~D.~M. Haldane {Phys. Lett.} {\bf 93A} (1983) 464; {Phys.
Rev. Lett.} {\bf 50} (1983) 1153; {J. Appl. Phys.} {\bf 57} (1985) 3359.
\bibitem{QHE} H. Levine, S. Libby and A. Pruisken, {Phys. Rev. Lett.} {\bf 51} (1983) 1915.
\bibitem{kt} V.~J. Berezinskii, {J.E.T.P.} {\bf 32} (1971) 493; J.~M. 
Kosterlitz and D.~V.  Thouless, {J. Phys.} {\bf C6} (1973) 1181; J.~M. 
Kosterlitz, {J. Phys.} {\bf C7} (1974) 1046.
\bibitem{ZZ} A.~B. Zamolodchikov and Al.~B. Zamolodchikov, {Ann. Phys.} {\bf 120} (1979) 253.
\bibitem{PW} P.~B. Wiegmann, {Phys. Lett.} {\bf 152B} (1985) 209. 
A.~M. Polyakov and P.~B. Wiegmann, {Phys. Lett.} {\bf 131B} (1983) 121; 
\bibitem{FadTakd} V.~O. Tarasov, L.~A. Takhtadjan and L.~D. Fadeev, {Theor. Math. Fiz.} {\bf 57}
(1983) 163. 
\bibitem{AffHal} I. Affleck, {Phys.
Rev. Lett.} {\bf 56} (1986) 408; {J. Phys. Condens. Matter} {\bf 1} (1989) 
3047; F.~D.~M. Haldane (unpublished). 
\bibitem{gross} D.~J. Gross, {Nucl. Phys.} {\bf B132} (1978) 439.
\bibitem{WPW} W. Bietenholz, A. Pochinski and U.-J. Wiese, {Phys. Rev. Lett.} {\bf 75} (1995) 4524.
\bibitem{asorey} M. Asorey and F. Falceto, 
{Phys. Rev. Lett.} {\bf 80} (1998) 234.
\bibitem{orland} P. Orland, NBI preprint, NBI-HE-96-35 (1996), hep-th/9607134.
\bibitem{kmo} M. Kudinov, E. Moreno and P. Orland, in Proc. of NATO Adv. 
Research
Workshop, Zakopane, {\it New Developments in Quantum Field Theory}, ed.
P.~H. Damgaard and J. Jurkiewicz (1998) 315.
\bibitem{feynman} R.~P. Feynman, {Nucl. Phys.} {\bf B188} (1981)
479.
\bibitem{babelon2} O. Babelon and C.~M. Viallet, {Commun. Math. 
Phys.} {\bf 81} (1981) 515; {Phys. Lett.} {\bf 103B}
(1981) 45.
\bibitem{alex} A.~D. Alexandrov and V.~A. Zalgaller, {\it Intrinsic Geometry of
Surfaces}, AMS Translations of Mathematical Monographs, {\bf 15}, ed. J.~M. Danskin (1967).
\bibitem{WW} E.~T. Whittaker and G.~N. Watson, {\it A Course of
Modern Analysis}, Cambridge University Press.
\bibitem{singer} I.~M. Singer, {Physica Scripta} {\bf 24} (1980) 817; M.~F. 
Atiyah, N.~J. Hitchin and I.~M. Singer, {Proc.
R. Soc. Lond.} {\bf A362} (1978) 425.; P.~K. Mitter and C.~M. Viallet, {Commun. Math. 
Phys.} {\bf 79} (1981) 457; {Phys. Lett.} {\bf 85B}
(1979) 246; P.~K. Mitter, in Proc. of NATO Adv. Study 
Inst., Carg{\`e}se, Aug. 1979, Plenum 
Press, New York (1980); M.~S. Narasimhan
and T.~R. Ramadas, {Commun. Math. Phys.} {\bf 67} (1979) 
21; M. Asorey and 
P.~K. Mitter, {Commun. Math. Phys.} {\bf 80} (1981) 43.
\bibitem{frad and suss} E. Fradkin and L. Susskind, {Phys. Rev.} {\bf D17} (1978) 2637.
\bibitem{hasenbusch} M. Hasenbusch and K. Pinn, {J. Phys.} {\bf A30} (1997)
63.
\bibitem{whittaker} E.~T. Whittaker, {\it A Treatise on Analytical Dynamics 
of Particles and Rigid Bodies}, Cambridge
University Press (1937). Note: there is an error in
Whittaker's discussion of the spherical pendulum 
in that he asserts that there is a choice for $\nu_{+}$ (called
$\lambda$ in his notation) 
which is purely
imaginary.
\bibitem{weinstein} A. Weinstein, {Amer. Math. Monthly} {\bf 49}
(1942) 521.
\bibitem{instantons} V.~A. Fateev, I.~V. Frolov and A.~S. 
Schwarz, {Nucl. Phys.} {\bf B154} (1979) 1; B. Berg and M.
L\"{u}scher, {Commun. Math. Phys.} {\bf 69} (1979) 57.
\bibitem{orland1} P. Orland, to appear.
\end{thebibliography}
\end{document}